\documentclass{tlp}
\nocite{*}
\submitted{16 January 2004}
  \revised{17 March 2005}
  \accepted{10 May 2005}
\usepackage{latexsym}
\usepackage{amssymb}
\usepackage{epsfig}
\usepackage{pstricks,pst-node,pst-plot,epsf}

\newcommand\bcmdtab{\noindent\bgroup\tabcolsep=0pt%

\begin{tabular}{@{}p{10pc}@{}p{20pc}@{}}}

\newcommand\ecmdtab{\end{tabular}\egroup}

\newtheorem{defn}{Definition}
\newtheorem{prop}[defn]{Proposition}

\newtheorem{thrm}[defn]{Theorem}
\newtheorem{coro}[defn]{Corollary}
\newtheorem{lema}[defn]{Lemma}

\title[Parametric Verification of a Group Membership Algorithm]
{Parametric Verification of a Group Membership Algorithm}

\author[A. Bouajjani and A. Merceron]
{AHMED BOUAJJANI\\
{\sc Liafa}, University of Paris 7, Case 7014, 2 place Jussieu, 75251 Paris 5, France\\
{\tt email: abou@liafa.jussieu.fr}\\
\and AGATHE MERCERON\\
{\sc Liafa} and {\sc Esilv-Gi}, Technical University Leonard de Vinci, 
92916 Paris La D\'efense, France\\
{\tt email: Agathe.Merceron@liafa.jussieu.fr}
}

\jdate{}

\pubyear{}

\pagerange{\pageref{firstpage}--\pageref{lastpage}}

\doi{S1471068401001193}

\newtheorem{lemma}{Lemma}[section]

\begin{document}
\label{firstpage}
\maketitle

\begin{abstract} We address the  problem  of verifying  clique
avoidance  in the TTP protocol. TTP
allows several stations  embedded in a car
to communicate. It has   many mechanisms  to ensure robustness to faults.
In particular, it has an algorithm that allows a  station to
recognize itself as faulty and leave the communication. 
This algorithm must satisfy the crucial 'non-clique' property: 
it is impossible to have two or more disjoint groups of  stations
communicating exclusively with stations in their own group. 
 In this paper,
we propose an automatic 
 verification method for an arbitrary number of stations $N$
and a given number of faults $k$.  We give an abstraction that
allows to model the algorithm by means of unbounded 
(parametric) counter automata.
We have checked the non-clique property
on this model  in the case of one fault, using the ALV tool 
as well as the LASH tool.
\end{abstract}

 \begin{keywords}
    Formal verification, fault-tolerant protocols,
parametric counter automata, abstraction
  \end{keywords}

\section{Introduction}
\label{intro}

The verification of complex systems, especially of software systems,
requires the adoption of powerful methodologies based on 
combining, and sometimes iterating, several  analysis techniques.
A widely adopted approach consists in combining abstraction techniques 
with verification algorithms 
(e.g., model-checking, symbolic reachability analysis, 
see, e.g., \cite{GS97,DT99,SS99}).
In this approach, non-trivial abstraction steps are necessary 
 to construct faithful abstract models (typically finite-state models) 
on which the required properties can be automatically verified.
The abstraction steps can be extremely hard to carry out
depending on how restricted  the targeted class of abstract models is.
Indeed, many aspects in the behavior of complex software systems cannot 
(or can hardly) be captured using finite-state models. 
Among these aspects, we can mention, e.g., 
(1) the manipulation of variables and data-structures (counters, queues, arrays, etc.) 
ranging over infinite domains, (2) parameterization (e.g., sizes of the data structures, 
the number of components in the system, the rates of errors/faults/losses, etc.). 
For this reason, it is often needed to consider abstraction steps which 
yield infinite-state models corresponding to extended automata,
i.e., a finite-control automata supplied with unbounded data-structures
(e.g., timed automata, pushdown automata, counter automata,
FIFO-channel automata, finite-state transducers, etc.) \cite{DT99}.
Then, symbolic reachability analysis algorithms 
(see, e.g., 
\cite{CH78,BW94,BEM97,BH97,KMMPS97,BGL98,WB98,BJNT00,AAB00,AJ01a,AJ01b})
 can be applied on these (abstract) extended automata-based models in order
to verify the desired properties of the original (concrete) system.
Of course, abstraction steps remain non-trivial in general for complex systems,
even if infinite-state extended automata are used as abstract models.

In this paper, we consider verification problems concerning 
a protocol used in the automotive and aerospace industry. The protocol,
called TTP/C, was designed at  
the Technical University of Vienna in order to allow 
 communication between  several devices (micro-processors) 
embedded in a car or plane, whose function is to control the safe execution 
of different driving actions \cite{vienna1,vienna4}.  

The protocol involves many mechanisms  to ensure robustness to faults.
In particular, the protocol involves implicit and explicit mechanisms
which allow to discard devices (called stations) which are (supposed to be) faulty.
This mechanism 
must ensure the crucial property:
{\em  all active stations form one single group
of communicating stations,
i.e., it is impossible to have two (or more) disjoint groups of active stations
communicating exclusively with stations in their own group}.

Actually, the algorithm is very subtle and 
its verification is a real challenge
for formal and automatic verification methods.
Roughly, it is a parameterized algorithm for $N$ stations
arranged in a ring topology. Each of the stations broadcasts
a message to all stations when it is its turn to emit.
The turn of each station is determined by a fixed time schedule.
Stations maintain informations corresponding to their view of          
the global state of the system: a membership vector,
consisting of an array with a 
parametric size $N$,
 telling which stations are active.
Stations exchange their views of the system and this allows them to recognize
faulty stations.
Each time a station sends a message, it sends also
the result of a calculation which encodes  its membership vector.
Stations compare their membership vectors to those received from sending stations. If a receiver disagrees with the membership vector
of the sender, it counts the received message as incorrect.
If a station disagrees with a majority of stations 
(in the round since the last time the station has emitted), it considers itself
as faulty and leaves the active mode (it refrains from emitting and 
skips its turn).
Stations which are inactive can return later to the active mode
(details are given in the paper).
Besides the membership vector,  each station $s$ maintains 
two integer counters in order 
to count  in the last round 
(since the previous emission of the station $s$) 
(1) the number of stations which have emitted 
and from which $s$ has received a correct message 
with  membership vector equal to its own vector at that moment
(the stations may disagree later concerning some other emitting station),
and (2) the number of stations 
from which $s$ received an incorrect message 
 (the incorrect message may be due to  a transmission 
fault or to a different membership vector).
The information maintained by each station $s$ depends tightly
on its position in the ring relatively to the positions
of the faulty stations and relatively to the stations which agree/disagree 
with $s$ w.r.t. each fault.

The proof of correctness of the algorithm and its automatic verification 
are far from being straightforward, 
especially in the parametric case, i.e., for any number of stations,
and any number of faults. 

The first contribution of this paper is to  prove that the algorithm
stabilizes to a state where  all
membership vectors are equal after precisely two rounds 
from the occurrence of the last fault in any sequence of faults.
The proof is given for the general case where re-integrating stations 
are allowed. To guarantee stabilization after $k$-faults in the
case of re-integration, we propose an algorithm 
slightly different from the one presented in \cite{vienna1}, which  
guarantees stabilization in the case of $1$ fault only. 
The generalization to $k$ faults makes an assumption on the failure model (made explicit in section \ref{infDesc}) that may not be realistic for a particular kind of messages called N-frames \cite{vienna1}. 

Then, we address the problem of verifying automatically
the algorithm.
We prove that, for every fixed number of faults $k$,
it is possible to construct an  abstraction of 
the algorithm (parameterized by the number of stations $N$) by means of
a parametric counter automaton.
This result is surprising since
(1) it is not easy to abstract the information related to the topology of the system
(ordering between the stations in the ring),
and (2) each station (in the concrete algorithm) has local variables 
ranging over infinite domains (two counters and an array with parametric bounds).
The difficulty is to prove that it is possible to encode the information
needed by all stations by means of a finite number of counters.
Basically, this is done as follows:
(1) We observe that a sequence of faults induces a partition
of the set of active stations (classes correspond to stations having
the same membership vector)
which is built by successive refinements: 
Initially, all stations are in the same set, and the occurrence
of each fault has the effect of splitting the class containing the faulty station
into two subclasses (stations which recognizes the fault, and the other ones).
(2) We show that there is a partition of the ring into a finite number 
of regions (depending on the positions of the faulty stations) such that, 
to determine at any time whether a station of any class can emit,
it is enough to know how many stations in the different classes/zones
have emitted in the last two rounds.
This counting is delicate due to the splitting of the classes after each fault.

Finally, we show that, given a counter automaton modeling the algorithm,
the stabilization property (after 2 rounds following the last fault)
can be expressed as a constrained reachability property (in CTL with Presburger predicates)
which can be checked  using symbolic reachability analysis tools
for counter automata (e.g., ALV \cite{bultan2} or LASH \cite{LASH}).
We have experimented this approach in the case of one fault.
We have built a model for the algorithm in the language of ALV, 
and we have been able to verify automatically that it converges to a single
clique after precisely two rounds from the occurrence of the fault.
Actually, we have provided a refinement of the abstraction
given in the general case which allows to build a simpler
automaton. This refinement is based on properties specific to the $1$
fault case that has been checked automatically using ALV.

The paper is organized as follows. Section \ref{infDesc} presents 
the protocol. In Section \ref{kfaults}, we prove the 
crucial non-clique property for $n$ stations: 
the stations that are still active do have the same membership vector
at the end of the second round following fault $k$.
Considering the $1$ fault case, 
section \ref{clique} presents how to abstract the protocol
parameterized by the number of stations $n$ as an automaton with counters
that can be symbolically model checked. Section \ref{autk} generalizes
the approach for a given number of faults $k$. Section \ref{concl}
concludes the paper. 
A preliminary version of this paper has appeared in \cite{BM02}.

\section{Informal Description of the Protocol}
\label{infDesc}

\begin{figure}
\vspace{2.5cm}
\includegraphics{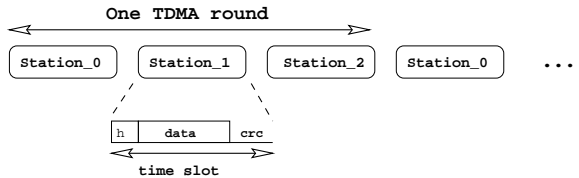}
\caption{A TDMA round for 3 stations.}
\label{tdmacycle}
\end{figure}

TTP is a time-triggered protocol. 
It  has a finite set $S$ of $N$ stations
and allows them 
 to communicate via a shared bus.\ Messages are broadcast to
all stations via the bus.\
Each station that participates in the communication sends a message
when it is  the {\em right} time to do so.\ Therefore, 
access to the bus is determined by a time division multiple
access (TDMA) schema controlled by the global
time generated by the protocol.\ A TDMA round is divided
 into {\em time slots}.\
The stations are statically 
ordered in a ring 
and time slots are allocated to the stations according to their
order.\
During its time slot, a station has exclusive message sending rights.\
A TDMA round for three stations is shown in Figure \ref{tdmacycle}.
When one round is completed, a next one takes place following the same
pattern.

TTP is a fault-tolerant protocol. Stations may fail
while other stations continue communicating with each other.
TTP provides different services to ensure robustness to faults, such as
 replication
of stations, replication of communication channels, 
bus guardian, fault-tolerant clock
synchronization algorithm, implicit acknowledgment, clique avoidance
mechanism,  \cite{vienna1,vienna4,bauer}. 
Several classes of faults are distinguished. \cite{steiner04}, for example,
focuses on faults that may appear at startup. In this paper, we focuse
on asymmetric faults. A symmetric fault occurs
when a station is {\em send faulty}, i.e., no other station can receive it properly,
or {\em receive faulty}, i.e., it cannot receive properly any message.
Asymmetric faults occur when an emitting station is received properly by more
than $1$ station, but less then all stations.  
We allow asymmetric faults to occur and consider symmetric faults as a special case of asymmetric faults.
For the protocol to work well, it is essential that (asymmetric)
faults do not
give rise to  cliques. 
In \cite{vienna1,vienna4}  {\em cliques} are understood
as  disjoint sets of stations communicating exclusively with each other.
In this paper, we focus on implicit acknowledgment and clique avoidance
mechanism, to be introduced shortly,
 and show that they prevent the formation of different
cliques,  clique is cast in its
graph theoretical meaning.

\subsection{Local Information}

When it is working or in
the {\tt active} state,
 a station sends messages in its time slot, 
listens to  messages broadcast by other stations and carries
local calculations. 
Each station $s$ stores locally  some information, in particular a 
{\em membership vector} $m_s$ and two counters, $C\!Acc_s$
and $C\!F\!ail_s$. 
A {\em membership
vector} is an array of booleans indexed
by $S$, the set formed by the $N$ stations.   It  indicates  the
stations that $s$ receives correctly (in a sense that will be
made precise below). If $s$ received
correctly the last message, also called {\em frame},
 sent by $s'$, then $m_s[s'] = 1$,
otherwise $m_s[s'] = 0$. A sending station is supposed to receive
itself properly, thus $m_s[s] = 1$ for a working station $s$.
The counters  $C\!Acc_s$
and $C\!F\!ail_s$ are used as follows.
When it is ready to send,
$s$ resets 
$C\!Acc_s$ and $C\!F\!ail_s$ to $0$. During the subsequent round,
$s$ increases $C\!Acc_s$ by $1$ each time it receives
a correct frame (this includes the frame it is sending itself)
 and it increases $C\!F\!ail_s$ by $1$ each time it receives
an incorrect frame. When no frame is sent (because the station that
should send is not working), 
neither $C\!F\!ail_s$ nor $C\!Acc_s$
are increased. 

\subsection{Implicit Acknowledgment}
\label{impAck}

Frames are broadcast over the bus to all stations but they are
not explicitly acknowledged. TTP has {\em implicit acknowledgment}. 
A frame is composed of a header, denoted by  {\tt h} in 
Figure \ref{tdmacycle},  a data field, denoted by {\tt data} and a CRC field
denoted by {\tt crc}.
The data field
contains the data, like sensor-recorded data,
 that a station wants to broadcast.
 The CRC field contains the
calculation of the Cyclic Redundancy Check done by the sending station. 
CRC is calculated
over the header, the data field and the individual membership
vector. When station $s$ is sending, it puts in the CRC field
the calculation it has done with its own membership
vector $m_s$.
Station $s'$ receiving a frame from station $s$
recognizes the frame as {\em valid} if all 
the fields have the expected lengths.
If the frame is valid, station $s'$ performs a CRC calculation
over the header and the data field it has just received,
and its own membership vector $m_{s'}$. 
It recognizes the frame as {\em correct} if it
has recognized it as valid
and its CRC calculation
agrees with the one put by $s$ in the CRC field. 
 Therefore, a correct CRC implies that sender $s$ and receiver $s'$
have the same membership vector.

{\em We also assume a converse:} if $s$ and $s'$ do not
have the same membership vector, the CRC is not correct.
This assumption may be strong and may not be met by 
special messages of TTP called N-frames. 
However, it  is valid for  X-frames and I-frames.

During normal operation, the sender $s$ performs two CRC checks 
over the frame received  from its first successor $s'$:

{\tt CheckIa} CRC calculation with $m_s[s] = 1$ and $m_s[s'] = 1$.

{\tt CheckIb} CRC calculation with $m_s[s] = 0$ and $m_s[s'] = 1$.

\noindent
If the CRC {\tt CheckIa} is correct, $s$ knows that $s$
and $s'$ have identical membership vectors, so it {\em implicitly} deduces
 that
its successor $s'$ has received its frame. Thus  $s$ assumes
that it is not faulty,  $m_s[s]$ remains $1$ and
{\tt CheckIb} is discarded. It increases $C\!Acc_s$
by one. One says that $s$ has reached its
{\em membership point}.

\noindent
If both {\tt CheckIa} and {\tt CheckIb}
fail, it is assumed that some transient disturbance has
corrupted the frame of  $s'$.  Thus
 $m_s[s'] $ is put to $0$ and  $C\!F\!ail_s$
is increased by $1$. The next station becomes the first successor of $s$.


\noindent
When  {\tt CheckIa}   fails but {\tt CheckIb} passes, either $s$
could be {\em send faulty}
or $s'$ could be {\em receive faulty}.
According to the confidence principle, $s$
assumes the latter,  puts the membership of $s'$ to $0$ and increases 
$C\!F\!ail_s$  by $1$. However,  $s$ 
 performs further  similar checks 
over the frame received from  the {\em next successor $s''$} for double
check:

{\tt CheckIIa} CRC calculation with $m_s[s] = 1$ and $m_s[s'] = 0$.

{\tt CheckIIb} CRC calculation with $m_s[s] = 0$ and $m_s[s'] = 1$.

\noindent
If {\tt CheckIIa} passes, $s$ is confirmed  that $s'$
is {\em received faulty}. It increases $C\!Acc_s$ 
  by $1$ (for $s''$)  and {\tt CheckIIb} is discarded. Again, $s$ 
has reached its
{\em membership point}.

\noindent
If {\tt CheckIIb} passes, $s$ assumes that it is itself  {\em send faulty}
 and $s'$ 
is non-faulty 
and leaves the active state.

\noindent
If both {\tt CheckIIa} and {\tt CheckIIb} fail,
$s$ considers that $s''$ is faulty. Therefore it puts $m_s[s''] = 0$,
it increases $C\!F\!ail_s$  by $1$ and will perform
{\tt CheckIIa} and {\tt CheckIIb} again with the next successor
following the same procedure.

It is  assumed that at least $3$ stations are active. 

\subsection{Clique Avoidance Mechanism}

The  {\em clique avoidance mechanism} reads as follows: 
Once per round, at the beginning of its time slot,
 a station $s$ checks whether $C\!Acc_s > C\!F\!ail_s$.
If it is the case, it resets both counters as already said above
and sends  a message.
 If it is not the case, the station fails.
It puts its own membership
vector bit to $0$, i.e., $m_s[s]=0$, and leaves the {\tt active} state,
thus will not send in the subsequent rounds.
The intuition behind this mechanism is that
 a station that fails to recognize a majority
of frames as correct, is most probably not working properly.
Other working stations, not receiving anything during the time slot
of $s$, put the bit of $s$ to $0$ in their own membership vector.

It should be noted that implicit acknowledgment and clique avoidance 
mechanism interfere 
with each other, which contributes to make the analysis of 
the algorithm difficult.

\subsection{Re-integration}

Faulty stations that have left the active state can re-integrate
the active state~\cite{vienna4,bauer}. 
The re-integration algorithm that we describe here differs slightly
from the one proposed in~\cite{vienna1} to guarantee 
stabilisation after $k$ faults. The algorithm proposed in~\cite{vienna1}
is enough to guarantee stabilisation after $1$ fault, 
but not  after $k$ faults.
A major difference  is that re-integration, with our algorithm, may last longer, since a station, once it has acquired
a membership vector, has to listen at least a full round before beeing able to send a frame, which is not necessarily the case with the algorithm
proposed in~\cite{vienna1}.

A re-integrating station $s$ copies the membership vector from
some active station. As soon as the integrating station  has a copy,
 it updates its membership vector
listening to the traffic following the same algorithm as other 
working stations.
During its first sending slot, it resets
 both counters, $C\!Acc_s$ and $C\!F\!ail_s$
to $0$, without sending any frame. During the following
round, it increases its counters and keeps updating its
membership vector as working stations do.
 At the beginning of its next sending slot,
$s$ checks whether $C\!Acc_s > C\!F\!ail_s$.
If it is the case, it puts $m_s[s]$ to $1$ and
 sends a frame, otherwise it leaves the {\tt active} state again.
Receiving stations, if they detect a valid frame, put the membership
of $s$ to $1$ and then  perform the CRC checks as described above.

\subsection{Example}


Consider a set $S$ composed of $4$ stations and suppose that all stations
 received correct frames from each other for a while.
This means that 
they all have identical membership vectors and $C\!F\!ail = 0$.
 After
station $s_3$ has sent, the membership vectors as well as  counters
$C\!Acc$ and $C\!F\!ail$  look as follows. 
Remember that there is no global resetting of $C\!Acc$ and $C\!F\!ail$.
Resetting is relative to the position of the sending station.\\


\begin{tabular}{c|cccccc}
\hline\hline
stations & $m[s_0]$  & $m[s_1]$  & $m[s_2]$  & $m[s_3]$  & $C\!Acc$ & $C\!F\!ail$ \\
\hline
$s_0$  & 1 & 1 & 1& 1 & 4 & 0 \\
$s_1$  & 1 & 1 & 1& 1  & 3 & 0\\
$s_2$  & 1 & 1 & 1& 1 & 2 & 0 \\
$s_3$  & 1 & 1 & 1& 1 & 1 & 0\\
\hline\hline
\end{tabular}

We suppose that a fault 
 occurs while $s_0$ is sending and that no subsequent fault occurs
for at least two rounds, calculated from the time
slot of $s_0$. 
We assume also that the frame sent by $s_0$ is recognized 
as correct by $s_2$ only. So the set $S$ is split in two subsets,
$S_1 = \{s_0, s_2\}$ and $S_0 = \{s_1, s_3\}$.

\begin{enumerate}
\item

Membership vectors  and counters after $s_0$ has sent: \\

\begin{tabular}{c|cccccc}
\hline\hline
stations & $m[s_0]$ & $m[s_1]$ & $m[s_2]$ & $m[s_3]$ & $C\!Acc$ & $C\!F\!ail$ \\
\hline
$s_0$  & 1 & 1 & 1& 1 & 1 & 0 \\
$s_1$  & 0 & 1 & 1& 1  & 3 & 1\\
$s_2$  & 1 & 1 & 1& 1 & 3 & 0 \\
$s_3$  & 0 & 1 & 1& 1 & 1 & 1\\
\hline\hline
\end{tabular}

\item

Membership vectors and counters after $s_1$ has sent. At this point  
{\tt CheckIa}   fails but {\tt CheckIb} passes for $s_0$.
However, {\tt CheckIa} passes for $s_3$ (because of the fault, both
{\tt CheckIa} and {\tt CheckIb} failed in the preceding time slot
for $s_3$).
Notice that $s_2$ does not have the same membership vector
as $s_1$,  so it does not recognize
the frame sent by $s_1$ as correct.\\

\begin{tabular}{c|cccccc}
\hline\hline
stations & $m[s_0]$ & $m[s_1]$ & $m[s_2]$ & $m[s_3]$ & $C\!Acc$ & $C\!F\!ail$ \\
\hline
$s_0$  & 1 & 0 & 1& 1 & 1 & 1 \\
$s_1$  & 0 & 1 & 1& 1  & 1 & 0\\
$s_2$  & 1 & 0 & 1& 1 & 3 & 1 \\
$s_3$  & 0 & 1 & 1& 1 & 2 & 1\\
\hline\hline
\end{tabular}

\item

Membership vectors  and counters after  $s_2$ has sent.
Now {\tt CheckIIa} passes for $s_0$, but  both 
{\tt CheckIa}  and {\tt CheckIb} fail for $s_1$ because its
membership vector differs with the one of $s_2$  on $s_0$.
 $s_3$ does not have the same membership vector
as $s_2$, so it does not recognize
the frame sent by $s_2$ as correct.
 \\

\begin{tabular}{c|cccccc}
\hline\hline
stations & $m[s_0]$ & $m[s_1]$ & $m[s_2]$ & $m[s_3]$ & $C\!Acc$ & $C\!F\!ail$ \\
\hline
$s_0$  & 1 & 0 & 1& 1 & 2 & 1 \\
$s_1$  & 0 & 1 & 0 & 1  & 1 & 1\\
$s_2$  & 1 & 0 & 1 & 1 & 1 & 0     \\
$s_3$  & 0 & 1 & 0& 1 & 2 & 2\\
\hline\hline
\end{tabular}

\item

Memberships and counters after the time slot of $s_3$, which cannot send due to the
clique avoidance  mechanism and
 leaves the {\tt active} state: \\

\begin{tabular}{c|cccccc}
\hline\hline
stations & $m[s_0]$ & $m[s_1]$ & $m[s_2]$ & $m[s_3]$ & $C\!Acc$ & $C\!F\!ail$ \\
\hline
$s_0$  & 1 & 0 & 1& 0 & 2 & 1 \\
$s_1$  & 0 & 1 & 0& 0  & 1 & 1\\
$s_2$  & 1 & 0 & 1 & 0 & 1 & 0     \\
$s_3$  & 0 & 0 & 0 & 0 & 0 & 0\\
\hline\hline
\end{tabular}

\item

Memberships and counters after  $s_0$ has sent again.
At this point {\tt CheckIa} succeeds for $s_2$ while
$s_1$ is still looking for a first successor. \\

\begin{tabular}{c|cccccc}
\hline\hline
stations & $m[s_0]$ & $m[s_1]$ & $m[s_2]$ & $m[s_3]$ & $C\!Acc$ & $C\!F\!ail$ \\
\hline
$s_0$  & 1 & 0 & 1& 0 & 1 & 0 \\
$s_1$  & 0 & 1 & 0& 0  & 1 & 2\\
$s_2$  & 1 & 0 & 1 & 0 & 2 & 0     \\
$s_3$  & 0 & 0 & 0 & 0 & 0 & 0\\
\hline\hline
\end{tabular}

\item

Memberships and counters  after the time slot of $s_1$, which
cannot send due the clique avoidance mechanism and
 leaves the {\tt active} state: \\

\begin{tabular}{c|cccccc}
\hline\hline
stations & $m[s_0]$ & $m[s_1]$ & $m[s_2]$ & $m[s_3]$ & $C\!Acc$ & $C\!F\!ail$ \\
\hline
$s_0$  & 1 & 0 & 1& 0 & 1 & 0 \\
$s_1$  &  0 & 0 & 0 & 0 & 0 & 0 \\
$s_2$  & 1 & 0 & 1 & 0 & 2 & 0     \\
$s_3$  &0 & 0 & 0 & 0 & 0 & 0 \\
\hline\hline
\end{tabular}

Membership vectors are coherent again at this point of time. 

\end{enumerate}

\section{Proving Clique Avoidance}
\label{kfaults}

In this section  we prove that if $k$ faults occur at a rate
of more than $1$ fault per two TDMA rounds and if no fault
occur during two rounds following fault $k$,
 then at the end of that second round,
all active stations have the same membership vector, so they form
a single {\em clique} in the graph theoretical sense.

Let us denote by $W$ the subset of $S$ that contains all
working stations.
We may  write $m_s = S'$ for a station $s$
with $S'\subseteq S$ as a short hand for $m_s[s'] = 1$ iff $s' \in S'$.
To prove coherence of  membership vectors
 we start with the following situation.
We suppose that stations of $W$ have identical membership vectors and all
have $C\!F\!ail_s = 0$. Because $m_s[s] = 1$ for any working station,
this implies that $m_s = W$ for any $s\in W$. 
Faults occur from this initial state.

 Let us define a graph as follows :  the nodes
are the stations, and there is an arc between
$s$ and $s'$ iff  $m_s[s'] = 1$. We recall that,
in graph theory, a {\em clique} is a complete subgraph, i.e., 
each pair of nodes is related by an arc. Thus initially, $W$ forms
a single clique in the graph theoretical sense.


\subsection{Introductory Example}
\label{introEx}

Let us illustrate how things might work  in the case of two faults. 
The first fault occurs when $s_0$ sends. We suppose
that only $s_1$ fails to receive correctly the frame sent by $s_0$.
 $S$ is split
as $S_1 =\{s_0, s_2, s_3\}$ and $S_0 = \{s_1\}$.
Membership vectors and counters after $s_0$ has sent.  At this point,
{\tt CheckIa}   passes for   $s_3$. \\

\begin{tabular}{c|cccccc}
\hline\hline
stations & $m[s_0]$ & $m[s_1]$ & $m[s_2]$ & $m[s_3]$ & $C\!Acc$ & $C\!F\!ail$ \\
\hline
$s_0$  & 1 & 1 & 1& 1 & 1 & 0 \\
$s_1$  & 0 & 1 & 1& 1  & 3 & 1\\
$s_2$  & 1 & 1 & 1& 1 & 3 & 0 \\
$s_3$  & 1 & 1 & 1& 1 & 2 & 0\\
\hline\hline
\end{tabular}\\
Notice that $\{s_0, s_1, s_2, s_3\}$, do not form a clique anymore,
the arc $(s_1, s_0)$ is missing. 

Membership vectors  and counters after $s_1$ has sent. At this point,
 {\tt CheckIa}   fails but {\tt CheckIb} passes for $s_0$.
$s_2$ and $s_3$ do not have the same membership vector as $s_1$,
so they don't accept its frame as correct.\\

\begin{tabular}{c|cccccc}
\hline\hline
stations & $m[s_0]$ & $m[s_1]$ & $m[s_2]$ & $m[s_3]$ & $C\!Acc$ & $C\!F\!ail$ \\
\hline
$s_0$  & 1 & 0 & 1& 1 & 1 & 1 \\
$s_1$  & 0 & 1 & 1& 1  & 1 & 0\\
$s_2$  & 1 & 0 & 1& 1 & 3 & 1 \\
$s_3$  & 1 & 0 & 1& 1 & 2 & 1\\
\hline\hline
\end{tabular}\\

Membership vectors and counters  after $s_2$ has sent. At this point, 
we suppose that a second fault
occurs. Neither $s_3$ nor $s_0$ recognize the frame sent by
$s_2$ as correct. $S_1$ is split in $S_{11} = \{s_2\}$
and $S_{10} = \{s_0, s_3\}$. At this point, 
$s_0$ keeps looking for a second  successor and both {\tt CheckIa}  and
 {\tt CheckIb}  fail for $s_1$.

\begin{tabular}{c|cccccc}
\hline\hline
stations & $m[s_0]$ & $m[s_1]$ & $m[s_2]$ & $m[s_3]$ & $C\!Acc$ & $C\!F\!ail$ \\
\hline
$s_0$  & 1 & 0 & 0& 1 & 1 & 2 \\
$s_1$  & 0 & 1 & 0& 1  & 1 & 1\\
$s_2$  & 1 & 0 & 1& 1 & 1 & 0 \\
$s_3$  & 1 & 0 & 0& 1 & 2 & 2\\
\hline\hline
\end{tabular}\\

Membership vectors and counters  after the time slot of $s_3$, which
is prevented from sending because of the clique avoidance mechanism:\\ 

\begin{tabular}{c|cccccc}
\hline\hline
stations & $m[s_0]$ & $m[s_1]$ & $m[s_2]$ & $m[s_3]$ & $C\!Acc$ & $C\!F\!ail$ \\
\hline
$s_0$  & 1 & 0 & 0& 0 & 1 & 2 \\
$s_1$  & 0 & 1 & 0& 0  & 1 & 1\\
$s_2$  & 1 & 0 & 1& 0 & 1 & 0 \\
$s_3$  & 0 & 0 & 0& 0 & 0 & 0\\
\hline\hline
\end{tabular}\\

One notices that $s_0$, then $s_1$
are prevented from sending because of the clique avoidance mechanism.
Membership vectors and counters  after the time slot of $s_1$:

\begin{tabular}{c|cccccc}
\hline\hline
stations & $m[s_0]$ & $m[s_1]$ & $m[s_2]$ & $m[s_3]$ & $C\!Acc$ & $C\!F\!ail$ \\
\hline
$s_0$  & 0 & 0 & 0& 0 & 0 & 0 \\
$s_1$  & 0 & 0 & 0& 0  & 0 & 0\\
$s_2$  & 0 & 0 & 1& 0 & 1 & 0 \\
$s_3$  & 0 & 0 & 0& 0 & 0 & 0\\
\hline\hline
\end{tabular}

 Coherence is achieved again after the time slot of $s_1$, where
$s_2$ remains the only active station. Though 
$S_{11}$ is smaller than $S_{10}$, the position of $s_2$ in the ring
as the first station of the round with the second fault allows it
to  capitalize
on  frames accepted in the round before and to win
over the set $S_{10}$.

\subsection{Proving a Single Clique after $k$ Faults}

The proof proceeds as follows. 
First we show a preliminary result.
If  $W$  is divided into
subsets $S_i$  in such a way that all stations in a subset
have the same membership vector, then stations inside a subset behave
similarly: if there is no fault, they recognize the same frames
as correct or as incorrect. Frames sent by stations from their own subset
are the ones that they recognize as correct, while frames sent
by stations from other subsets are all recognized as incorrect.

Then we show that the occurrence of faults does
produce such a partitioning, i.e., 
 after fault $k$, $W$ is divided into
subsets $S_w$,  
where $w \in \{0, 1\}^k$.
Indeed, as illustrated by the example above,
after $1$ fault, $W$ is split in $S_1$, the stations that recognize
the frame as correct, and $S_0$, the stations that do not
recognize the frame as correct. Because any station recognizes itself
as correct, $S_1$ is not empty.
Now, suppose that a second fault occurs. 
Assume that the second fault occurs when a station from set $S_1$ sends.
As before,  set $S_1$ splits into $S_{11}$, the stations
that recognize the frame as correct, and $S_{10}$, the stations
that do not recognize the frame as correct.
Again $S_{11}$
is not empty. Set $S_0$ becomes $S_{00}$ because
stations in $S_0$ don't have the same membership vector
as stations in $S_1$.  And the process
generalizes. If a station $s$ from a set $S_w$ sends when fault $k$ occurs,
$S_w$ splits into $S_{w1}$ and $S_{w0}$ with $S_{w1}\neq \emptyset$.
Then, we show  that two stations  $s$ and $s'$ have the same
membership vector if and only if they belong to the same set $S_w$.
Using the preliminary lemma,
we have  a result about
the incrementation of  the counters $C\!Acc$ and $C\!F\!ail$,
namely, all stations  from a set $S_w$ increment $C\!Acc$
if  a station from $S_w$ sends, and increment $C\!F\!ail$
if a station from $S_{w'}$ sends, where $w\neq w'$.
From this, we can deduce our main result:
 in the second round after  fault $k$,
only stations from a single set $S_w$ can send. It follows that,
at the end of that second round, there can be at most only one clique.

First we give a lemma that says that, 
if two stations have the same membership vector,
then they recognize mutually their frames as correct. 

\begin{lema}
\label{sameMMBship}
Let   $s $ and $ s' \in W$ with $m_s =  m_{s'}$. 
Then, 
$m_s[s'] = 1$ and $m_{s'}[s] = 1$.
\end{lema}

\begin{proof}
$s, s'\in W$ means $m_s[s] = 1$ and $m_{s'}[s'] = 1$.
Because $s$ and $s'$ have the same membership vector, one has
$m_s[s'] = 1$ and $m_{s'}[s] = 1$.
\end{proof}

Now we give our preliminary result when active stations are divided
into subsets $S_i$  in such a way that all stations in a subset
have the same membership vector.

\begin{prop}
\label{counters}
Suppose that $W$ is divided into $m$ subsets $S_1, \ldots , S_m$
such that $s $ and $ s' $
have the same membership vector iff 
$s$ and $s'$ belong to the same subset $S_i$, $1 \leq i\leq m$.
Let $s\in S_i$, $1 \leq i\leq m$.
Assume  that no fault occurs.
Then, each time some other station $s'$ is sending:
\begin{enumerate}
\item
 if  $s'\in S_i$, 
 $s$ increases $C\!Acc_s$ by $1$ and keeps the membership 
bit of $s'$ to $1$,

\item
If $s' \in S_{j}$, $j\neq i$, 
\begin{enumerate}
\item
either $s$ increases  $C\!F\!ail_{s}$
by $1$ and puts the membership 
bit of $s'$ to $0$,
\item
or $s$ leaves the active state.
\end{enumerate} 
\end{enumerate} 
\end{prop}

\begin{proof}
We prove this proposition for the general case where integrating stations are
allowed.

\noindent
Let $s\in S_i$, or, 
if $s$ is a station about to integrate, suppose
it has copied its membership vector
from an active station belonging to  $S_i$.

\noindent
First suppose $s'\in S_i$ or, 
if $s'$ is an integrating station,  $s'$ has copied 
its membership vector from some active station $s'' \in S_i$. 
Following the integration policy, both $s$ and $s'$ put the membership bit of $s'$ to $1$ when $s'$ sends.
Because $s$ and $s'$ have the same
membership vector 
and because no other fault occurs,
$s$ recognizes as correct the frame sent by $s'$.
So it increases $C\!Acc_s$ by $1$ when $s'$ sends and keeps the membership 
bit of $s'$ to $1$, also if $s$ is performing {\tt CheckIa} or
 {\tt CheckIIa}.

\noindent
Suppose now $s' \in S_j$, $j\neq i$.
As above, if $s'$ is an integrating station,  $s'$ has copied 
its membership vector from some active station $s'' \in S_j$.
Because $s$ and $s'$ do not have the same
membership vector,
$s$ does not recognize the frame sent by $s'$ as correct.
If $s$ has reached its membership point
already or is a station about to integrate, 
it increases $C\!F\!ail_s$ by $1$  and 
puts the membership 
bit of $s'$ to $0$.

\noindent
If $s$ has not reached its membership point yet, it performs
either {\tt checkI} or {\tt CheckII}.
Suppose first $s$ performs {\tt CheckIa} and {\tt CheckIb}, 
i.e., $s'$ could be  the first successor of $s$. 
Because $s'\in S_j$, {\tt CheckIa} does not pass.
Hence, $s$ increases $C\!F\!ail_s$ by $1$  and 
puts the membership 
bit of $s'$ to $0$.

\noindent
Suppose now  $s$ performs {\tt CheckIIa} and {\tt CheckIIb},
i.e.,  $s'$ could be the second successor of $s$. If $s$ and $s'$ disagree
only on the bit for $s$, then {\tt CheckIIb} fails for $s$
which leaves the active state. Otherwise, $s$ simply does not recognize
as correct the frame sent by $s'$; 
it increases $C\!F\!ail_s$ by $1$,
puts the membership 
bit of $s'$ to $0$ (and continues looking for a second successor).
\end{proof}

Now we prove a crucial proposition. Namely, occurence of faults
divide the active stations into subsets 
characterized by their membership vectors.

\begin{prop}
\label{justafters_kInt}
At the end of the time slot of $s^k$, 
the station where fault $k$ occurs 
$k\geq 1$,
$W$ is partitioned into subsets $S_w$, with
 $w \in \{0, 1\}^k$, such that 
two stations $s \in S_w$ and $ s' \in S_{w'}$
have the same membership vector iff
$w = w'$.
\end{prop}

\begin{proof}
We proceed by induction on $k$.

\noindent
Basis : $k = 1$. 
Let $s^1$ be the station which is sending
when the first fault occurs.
Before the time slot of $s^1$, all active stations, plus $s^1$
if $s^1$ is an integrating station,  have
the same membership vector, namely $W$, by hypothesis on the initial state.
In case  $s^1$ is an integrating station, all stations put the  membership
bit of $s^1$ to $1$ by the integrating station policy.
 We denote by $S_0$  the subset of $W$ 
that {\em failed} to  receive correctly the frame sent by $s^1$
while we denote by $S_1$  the subset of stations 
that {\em accepted} the frame as  correct.

\noindent
All stations in $S_1$ could receive correctly the frame sent by $s^1$,
consequently their membership vectors do not change.
None of the station in $S_0$ could receive correctly the frame
sent by $s^1$ because of the fault. 
Hence they all put the membership bit
of $s^1$ to {\em false}, also in the case
where $s^1$ was an integrating station. Thus they all have the same
membership vector, namely $W \setminus \{s^1\}$. 

Suppose the result is true till fault $k-1$.

\medskip
Induction step : we prove that the result is true for $k$, $k > 1$.
Let $s^k$ be the station which is sending when fault $k$ occurs.
By induction hypothesis, at the end of the time slot of $s^{k-1}$,
$W$ is partitioned into sets $S_w$ with $w\in \{0 | 1\}^{k-1}$.
 There is a unique $w \in \{0 | 1\}^{k-1}$ with $s^k \in S_w$
or such that the membership vector of $s^k$ is identical to 
stations of $S_w$, in case $s^k$ is an integrating station.

\noindent
First, we show that $S_w$ splits into $S_{w0}$ and $S_{w1}$.
By Lemma~\ref{sameMMBship},  $m_s[s^k] = 1$ for any station $s\in S_w$.
Let us denote by $S_{w1}$ the subset of $S_w$ that could receive
$s^k$ correctly and by $S_{w0}$ the subset of $S_w$ that could 
not receive
$s^k$ correctly. 
Obviously, $S_{w1}$ and $S_{w0}$ partition $S_{w}$.
After the time slot of $s^k$,
all stations in $S_{w1}$ keep the membership of
$s^k$ to $1$, while all stations in $S_{w0}$ put it to $0$.
Thus $s, s' \in S_w$ have the same membership vector if and only
if $s, s' \in S_{w1}$ or $s, s' \in S_{w0}$.\\
Consider some $w'  \in \{0 | 1\}^{k-1}$, with $w'\neq w$ and 
$S_{w'} \neq\emptyset$. Let $s'\in S_{w'}$. We show that
$S_{w'}$ becomes $S_{w'0}$ and that the condition on the
membership vectors still holds.\\
$w'\neq w$ means that the $m_{s'}$ and  $m_{s^k}$ differ
on some bit $s^\ast$.
  So, 
  $s'$ cannot  recognize as correct
the frame sent by $s^k$ and $S_{w'}$ can be denoted by $S_{w'0}$.\\
Obviously, $\forall s \in S_{w1}: m_{s'} \neq m_{s}$.
Could it be that now stations $S_{w'0}$ and $S_{w0}$ have the
same membership vector? This could be only if their membership
vectors differed only on the bit for $s^k$. But this would mean
that station $s^k$ has already emitted,
that a fault occured  and station in $S_{w'}$ 
did not accept the frame as correct
while stations in  $S_{w}$ did. Because all stations emit in one round,
some station $u$ 
from  $S_{w'}$ has emitted. By Proposition \ref{counters},
membership vectors of stations in $S_{w'}$ and $S_{w}$ differ on $u$
and $u\neq s^k$.
Hence, stations $S_{w'0}$ and $S_{w0}$ have different
 memberships vectors.
\noindent
Using a similar argument, membership vectors
of stations in $S_{w''0}$ and $S_{w'0}$ remain different
with  $w''  \in \{0 | 1\}^{k-1}$, $w'\neq w'' \neq w$
and  $S_{w''} \neq\emptyset$.
\end{proof}

Finally, we show that only stations from a unique set $S_w$ are
able to send in the second round following the first fault.

\begin{thrm}
\label{genP1}
Suppose some station is able to send in the second round following 
 fault $k$.
Let us denote this station by $s$.
By Proposition \ref{justafters_kInt}, exists some 
$w  \in \{0 | 1\}^{k}$ such that $s \in S_w$.
Then, only stations from  $S_w$ can send in the second round
following fault $k$.
\end{thrm}

\begin{proof}
Let $C\!Acc_s$ and $C\!F\!ail_{s}$ when $s$ performs the
clique avoidance mechanism. 
By Proposition~\ref{counters},  
$C\!Acc_s = \mid \!\{s'\in S_w$ s. t. $ s'$ sent in the 
first round following fault $k \}\!\mid$ and
$C\!F\!ail_{s} = \Sigma_{w'\neq w}\mid \!\{s'\in S_{w'}$ s. t. $ s<s',$
and $ s'$ sent in the first round following fault $k \}\!\mid$,
where $<$ refers to the statical order among stations.  
Because $s$ is able to send, one has $C\!Acc_s > C\!F\!ail_{s}$ at the 
beginning of the time slot of $s$ in the second round. 

\noindent
Let $t$ be the follower station of $s$ in the statical order
ready to send after $s$. Is $t$
able to send, or is it prevented from sending
by the clique avoidance mechanism? We show that $t$ is able
to send if and only if it belongs  to $S_w$.

\noindent
Suppose $t\in S_w$. 
When its time slot comes in the second round, it has increased
$C\!Acc_t$ by $1$ when $s$ has sent in the second round.
Because, in the first round following fault $k$,
 $t$ increases its counters as $s$ does 
by Proposition~\ref{counters}, one has 
$C\!Acc_t =C\!Acc_s$, or $C\!Acc_t =C\!Acc_s +1$
in case $s$ was an integrating station,  and  $C\!F\!ail_t = C\!F\!ail_s$
at the beginning of the time slot of $t$ in the second round.
 Thus $C\!Acc_t > C\!F\!ail_t$ and $t$ is able to send as well.

\noindent
Suppose now $t\in S_{w'}$ for some $w'\neq w$.
$C\!Acc_t =  \mid \!\{s'\in S_{w'} \mid  t\leq s' \leq s^k$
and $s'$ sent in the first round following fault $k \}\!\mid$.
Indeed, between $s^k$ and its present time slot, $t$ has not accepted 
any frame since only $s$ has sent.
However, by Proposition~\ref{counters}, $C\!Acc_t \leq  C\!F\!ail_s$ since
all frames accepted by $t$ are not recognized as correct by $s$.
For a similar reason,
 $C\!Acc_s \leq  C\!F\!ail_t$. 
 It follows that, at the beginning of its time slot,
  $C\!Acc_t < C\!F\!ail_t$
and $t$ is prevented from sending.

\noindent
A similar argument  can be repeated to all stations 
ready to send in the second round  giving the result.
\end{proof}

From Theorem~\ref{genP1}, one deduces
that, at the end of the second round following fault $k$,
for any station $s\in S_w$:  $m_s = S_w$.
Using Lemma~\ref{sameMMBship}, this gives our safety property about cliques. 

\begin{coro}
\label{safeClique}
At the end of the second round following fault $1$, all 
working stations form a single clique in the graph theoretical sense. 
\end{coro}

\section{Automatic Verification: the  $1$ Fault Case}
\label{clique}

In the case of a single fault, the set $W$ of active stations
is divided into two subsets, $S_1$ and $S_0$.
The set $S_1$ is not empty as it contains $s^1$, the station
that was sending when the fault occurs.
We assume that no other fault occurs for the next two
rounds, 
 a round is taken with the beginning of the time
slot of $s^1$. We want to prove automatically for an arbitrary number $N$ of
stations
that, at the end of the second round following the fault,
all working stations form a single non-empty clique.
To achieve this goal, we need a formalism to model the protocol
and a formalism to specify the properties that the protocol must satisfy.
To model the protocol, we take synchronous automata extended 
 with parameters and
counters. To specify the properties, we take the temporal logic CTL.
To keep the number of parameters as low as possible, we do not consider
re-integrating stations.

To be able to verify automatically the protocol for
the parametric case where the number of stations 
is a paramater $N$, we need to abstract the behavioural model
of the $N$ identical extended automata into a single extended
automaton. The abstraction we use is the standard 
(infinite) counter abstraction
and can be automatized. The novelty and difficulty of our
case study, compared with other examples
using a similar abstraction technique \cite{pnuelicav02,delzano},
 lies in the fact that each individual extended automaton
that models one station has local infinite variables : two counters, 
$C\!Acc$ and  $C\!F\!ail$, and a 
membership vector $m$ that all depends on the paramater $N$.
Applying counter abstraction directly would lead to a too coarse model,
useless for verification. Consequently, before we apply the abstraction,
we perform a  transformation of the $N$ extended automata in order
to replace local variables in guards by a finite number of global counters.
The successive models we obtain are related by a (bi)simulation
property. 
  
We divide the presentation in four main parts:  
first, we draw straightforward consequences from Section  \ref{kfaults}
for the $1$ fault case. 
 Second, we give the behavioural model   of the protocol under the form
of  $N$
synchronous automata with local  variables. 
Third we show how we build the abstract model  
replacing local  variables by a finite number of 
global counters, strenthening guards
and performing the usual counter abstraction
 in such a way that each  successive model (bi)simulates the previous one.
  Finally we give the properties that have been automatically proved 
on the resulting model which consists
of a single extended automaton. This establishes
 the 'non-clique' property.

\subsection{Properties of the $1$-fault case}

We will make use of the results presented in this section 
in the stepwise transformations 
 where local variables are replaced by a finite number
of global counters.

\begin{prop}
\label{leave-active}
Let $s$ be the sending station when the fault occurs. In the  round
following the fault, 
the $3$ conditions below are equivalent :
\begin{enumerate}
\item
$s$ leaves the active state after {\tt CheckI} and {\tt CheckII},
\item
the  two follower stations of $s$ have $1$ everywhere
in their membership vector, except for the bit for  $s$
which is $0$,
\item
the two follower stations of $s$ are in $S_0$.
\end{enumerate}
\end{prop}
\begin{proof}
A simple consequence of the fact that before the fault, all membership vectors
are all equal with $1$ in each bit.
\end{proof}

\begin{prop}
\label{leave-active-1}
Let $s$ be the sending station when the fault occurs and 
suppose that no fault occurs
during the two subsequent rounds.
Then, only station $s$ can leave the active state because 
of {\tt CheckI} and {\tt CheckII} in the round following the fault.
In later rounds, {\tt CheckI} and {\tt CheckII} do not play
any r\^ole.
\end{prop}
\begin{proof}
If $s$ has left because of {\tt CheckI} and {\tt CheckII}, 
there is nothing to prove.
Otherwise, the first or second successor of  $s$ belongs
to $S_1$, so either {\tt CheckIa} or {\tt CheckIIa}  succeeds
in this round or in later rounds.
Consider now  $s'\neq s$.
Could $s'$ leave the active state because of {\tt CheckI} and {\tt CheckII}~?
Consider $s''$ the station sending after $s'$.
Suppose that {\tt CheckIa} fails (otherwise there is nothing to prove).
This means that $s'$ and $s''$ have different membership
vectors. Because no new fault occurs, the difference in the 
membership vector can not be at the bit for $s'$ only.
Thus $s'$ discards both {\tt CheckIa} and {\tt CheckIb} and keeps
looking for a first successor. 
\end{proof}

\begin{figure}
\centering\epsfig{file=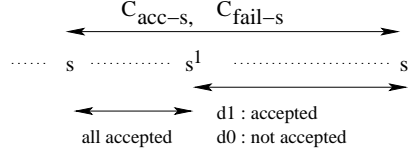, height=2cm}
\caption{Illustrating proof of Theorem \ref{countingAcc}.}
\label{LemCnt}
\end{figure}

The result below shows that we  can replace the $N$ individual 
counters $C\!Acc$ and  $C\!F\!ail$ by two global counters  $d0$ and $d1$.
Let  $d1$ be a counter  to count how many stations of $S_1$
have sent so far in the round since the fault occurred. 
Let $d0$ be a similar counter for $S_0$.
Let $s$ be a station ready to send. Assume $s\in S_1$.
How much is $C\!F\!ail_{s}$ ? It is exactly given by $d0$.
 How much is $C\!Acc_s$ ? Generally, it is more than $d1$.
One has to add all stations that have emitted before the fault
since the last time slot of $s$, because
$s$ has recognized them all as correct, see Figure \ref{LemCnt}.
However, this number can be calculated exactly with the
help of $d1$ and $d0$ only as  
Theorem \ref{countingAcc} shows. 

\begin{thrm}
\label{countingAcc}
Let $s$ a station ready to send and suppose there is only $1$ fault..

 In the round following  fault
$1$:
\begin{enumerate}
\item 
If  $s \in S_1$, then $C\!Acc_s = \mid\! W\!\mid - d0$ 
and $C\!F\!ail_{s} = d0$.
\item
If  $s \in S_0$, then $C\!Acc_s = \mid\! W\!\mid - d1$ and
$C\!F\!ail_{s} = d1$.
\end{enumerate}

In later rounds:
\begin{enumerate}
\item 
If  $s \in S_1$, then $C\!Acc_s = \mid\! S_1\!\mid$ and
$C\!F\!ail_{s} = \mid\! S_0\!\mid$.
\item
If  $s \in S_0$, then $C\!Acc_s = \mid\! S_0\!\mid$ and
$C\!F\!ail_{s} = \mid\! S_1\!\mid$.
\end{enumerate}
\end{thrm}
\begin{proof}
First round after the fault.
We prove the first item only, the second one being dual.
Since there was no fault in the last round, $s$ has recognized as 
correct all stations between itself and $s^1$ in the last round.
By Proposition \ref{counters},
 it has recognized
as correct all stations of $S_1$ that have sent so far in the round.
This is illustrated in Figure \ref{LemCnt}.
Thus $C\!Acc_s = \mid\!\{ s' \mid s\leq s' < s^1\}\cap W\!\mid + d1$, where
$\leq$ and $<$ refer  the statical order among stations.
$\{ s' \mid s\leq s' < s^1\}\cap W = 
W \setminus (\{ s' \mid s^1\leq s' < s\} \cap W)$.
Thus 
$\mid\! \{ s' \mid s\leq s' < s^1\}\cap W\!\mid = \mid\! W\!\mid -d1-d0$.
This gives $C\!Acc_s = \mid\! W\!\mid -d0$.
By  Proposition~\ref{counters},  $C\!F\!ail_{s} = d0$.

Second and later rounds after the fault.
Let $s \in S_1$ be the  station whose time slot comes first
in the second round.
This station is the faulty station itself. By Proposition~\ref{counters},
$C\!Acc_s$ is the number of stations of $S_1$ that could send in the previous
round, i.e., that are still  active, which is precisely $\mid\! S_1\!\mid$.
Similarly, $C\!F\!ail_{s} = \mid\! S_0\!\mid$. If $s$ can send,
$\mid\! S_1\!\mid$ remains unchanged, if $s$ cannot send,
$\mid\! S_1\!\mid$ diminishes by $1$. At the beginning
of the follower time slot, $\mid\! S_1\!\mid$ is exactly the number of stations
from $S_1$ that have send in the round preceding this time slot, and
similarly for $S_0$. Let us call $s'$ the station corresponding to that
time slot. If $s' \in S_1$, using Propositions~\ref{counters},
 $C\!Acc_{s'} = \mid\! S_1\!\mid$
and $C\!F\!ail_{s} = \mid\! S_0\!\mid$. 
If $s' \in S_0$, the dual is true. If $s' \not\in S_0\cup S_1$,
$s'$ has not send in the preceding round. In any case, at the end of the time
slot of $s'$,
 $\mid\! S_1\!\mid$ 
is exactly the number of stations
from $S_1$ that have send in the round, and
similarly for $S_0$. Using Proposition~\ref{counters} and repeating
the same argument for all 
 time slots that follow gives the result.
\end{proof}

\subsection{Behavioural model}

We give in this section a formal description of the TTP membership algorithm.
For that, we consider a general specification formalism for parametrized networks.
We assume that systems can have an arbitrary number $N$ of components
which may share global variables and can have also local variables.
Moreover, these components can communicate by broadcasting messages.
To describe the behaviour of each of the components, we adopt an extended automata-based 
formalism (described using the notations of~\cite{pnueli}) 
where each transition between control state 
is a guarded command which may involve (broadcast) communications.
We assume that the executions of all the parallel components are synchronous following
the semantical model of~\cite{ieee}
(i.e., all components have the same speed w.r.t. a global
logical clock defining a notion of execution step in the system,
and at each step, all the operations, including broadcast communications, are instantaneous).

The formal semantics of such models can be defined in terms of a transition system.
A state defines a global configuration of the network 
corresponding to the values of the global variables together
with the values of all local variables (including the control states) 
for each of the $N$ components of the network.
Transitions between states are defined straightforwardly according to the
semantical model mentioned above. The behaviours of the network
are defined as the possible execution paths in the so defined transition system.

Before giving the formal description of the TTP, let us introduce some notations.
We denote the sending of a broadcast message $a$ by $a!!$
and the receiving of a broadcast message $a$ by $a??$.
Otherwise, $\lfloor\!\rceil$ denotes non-deterministic choice, 
 $t^{\oplus\oplus}$ stands for $t:=t\oplus 1$, i.e.,
the value of $t$ in the next state is incremented by $1$  modulo $N$,  $A^{++}$ stands for $A:=A + 1$,  and '{\tt ;}'
denotes sequentiality. Assignments separated by '{\tt ,}' on
the right side of  $\longrightarrow $ could happen in any order.
 Variables not mentioned in 
transitions remain unchanged. 

Figure~\ref{pg1} gives the formal specification of the TTP membership algorithm.

\begin{figure}
\noindent
$
\begin{array}{lcllcll}
$in$\ N & :& $int,$ N > 1 & \ \ \ $in$\ fault & : & $boolean\ signal$ \\
\end{array}
$

\medskip

$\parallel_{i=1}^N P[i]$, \ \ \ 
where $P[i]$ is the following program:

\bigskip

%
$
\begin{array}{lcllcl}
\ A[i]& : & $int, init $N+1 -i & F[i] & : & $int, init $0\\
\ m[i]& :& $array of boolean, init $1 
    & t[i] & : & $int, init $1 \ \ // $\ t for turn$\\
\  s[i] & : & $int, init $0  // $\ identity of the faulty station$\\
\end{array}
$ \\

loop forever do \ \ \ \ \ \ \ \ \ \ \ \ \ \ \ \ \ \ \ \ \ \ \ \ \ \ \ \ \ \ \ \ \ \ \ \ \ \ \ \ \ \ \ \ \ \ \ \ \ \ \ \ \ \ \ \ \ \ \ \ \ \ \ \ \ \ \ \ \ \ \ \ \ \ \ \ \ \ \ \ \ \ \ \ \ \ \ \ \ \ \ \ \  \\
$
\begin{array}{ll}
 \   \   l_{in} & \left \{ \begin{array}{l}
                    \lfloor\!\rceil\ t[i] = i \ \wedge\  \neg fault \\
         \ \ \ \longrightarrow \
         emit!!, t[i]^{\oplus\oplus}, A[i]:=1;\ l_{in} \\
	 \lfloor\!\rceil\ t[i] = i \ \wedge\  fault \\ 
          \ \ \  \longrightarrow \
	 emit!!, t[i]^{\oplus\oplus}, A[i]:=1, s[i]:=t[i];\ l_1 \\
	 \lfloor\!\rceil\ t[i] \neq i \ \wedge\ emit??\ \wedge\  \neg fault\\ 
	 \ \ \ \longrightarrow\
          t[i]^{\oplus\oplus}, A[i]^{++};\ l_{in} \\
	  \lfloor\!\rceil\ t[i] \neq i \ \wedge\  emit?? \ \wedge\  fault\\ 
	  \ \ \  \longrightarrow\
           \ t[i]^{\oplus\oplus}, A[i]^{++}, s[i]:=t[i];\ l_1 \\
	  \lfloor\!\rceil\ t[i] \neq i \ \wedge\  emit?? \ \wedge\  fault \\  
	  \ \ \ \longrightarrow \
          t[i]^{\oplus\oplus}, F[i]^{++}, s[i]:=t[i], 
	  m[i][t[i]]:=0;\ l_0
	  \end{array} \right. \\
\\
\   \    l_0 & \left \{ \begin{array}{l}
     \lfloor\!\rceil\ t[i] = i \ \wedge\  A[i] > F[i]\\
       \ \ \ \longrightarrow \ 
         emit!!, t[i]^{\oplus\oplus}, A[i]:=1, F[i]:=0;\ l_0 \\
	 \lfloor\!\rceil\ t[i] = i \ \wedge\  A[i] \leq F[i]\\
        \ \ \ \longrightarrow \ 
          \neg emit!!, t[i]^{\oplus\oplus};\ l_F \\
	  \lfloor\!\rceil\ t[i] \neq i \ \wedge\  emit??  \ \wedge\  m[t[i]] = m[i]\\
	  \ \ \   \longrightarrow \ t[i]^{\oplus\oplus}, A[i]^{++};\ l_0\\
	  \lfloor\!\rceil\	 t[i] \neq i \ \wedge\  emit?? \ \wedge\  
	  m[t[i]] \neq m[i] \\ 
        \ \ \ \longrightarrow \ 
          t[i]^{\oplus\oplus}, F[i]^{++}, m[i][t[i]]:=0;\ l_0 \\
      \lfloor\!\rceil\ t[i] \neq i \ \wedge\  \neg emit??  \\ 
      \ \ \ \longrightarrow \ t[i]^{\oplus\oplus},  m[i][t[i]]:=0;\ l_0
	\end{array} \right. \\
\\
\  \    l_1 & \left \{ \begin{array}{l}
     \lfloor\!\rceil\ t[i] = i \ \wedge\  A[i] > F[i] \\
     \ \ \ \longrightarrow \ 
         emit!!, t[i]^{\oplus\oplus}, A[i]:=1, F[i]:=0;\ l_1 \\
	 \lfloor\!\rceil\ t[i] = i \ \wedge\  A[i] \leq F[i] \\
	 \ \ \ \longrightarrow \ 
	 \neg emit!!,
          t[i]^{\oplus\oplus};\ l_F \\
	  \lfloor\!\rceil\ t[i] \neq i \ \ \wedge\  emit?? \ \wedge\ i=s[i]
	  \ \wedge\ t[i]=s[i]+2
	  \ \wedge\ \\
          \ \ \ \    (m[t[i]][s[i]]=  m[t[i]\! -\! 1][s[i]] = 0)\\
	   \ \ \ \longrightarrow \ 
          t[i]^{\oplus\oplus};\ l_F \\
	  \lfloor\!\rceil\ t[i] \neq i \ \ \wedge\  emit??  \ 
	  \wedge\ i=s[i]\ \wedge\ t[i]=s[i]+2
	   \wedge \\
     \ \ \ \    \neg (m[t[i]][s[i]]= m[t[i]\! -\! 1][s[i]]=0) \\
	    \ \ \ \longrightarrow  
          t[i]^{\oplus\oplus},  A[i]^{++};\ l_1\\
	  \lfloor\!\rceil\ t[i] \neq i \ \wedge\  emit??  \ \wedge\ (i\neq s[i]
	  \vee t[i]\neq s[i]+2) \wedge  m[t[i]] = m[i] \\
	    \ \ \ \longrightarrow \ t[i]^{\oplus\oplus}, A[i]^{++};\ l_1\\
	  \lfloor\!\rceil\	 t[i] \neq i \ \wedge\  emit?? \ \wedge\ (i\neq s[i]
	  \vee t[i]\neq s[i]+2) \wedge 
	  m[t[i]] \neq m[i] \\  
	  \ \ \ \longrightarrow \ 
          t[i]^{\oplus\oplus}, F[i]^{++}, m[i][t[i]]:=0;\ l_1 \\
      \lfloor\!\rceil\ t[i] \neq i \ \wedge\  \neg emit??  \\  
      \ \ \ \longrightarrow \ t[i]^{\oplus\oplus},  m[i][t[i]]:=0;\ l_1
	\end{array} \right. \\
\\
 \  \  l_F & \left \{ \begin{array}{l}
       \lfloor\!\rceil\ t[i] = i \\ 
       \ \ \ \longrightarrow \ 
         \neg emit!!, t[i]^{\oplus\oplus};\ l_F\\
	\lfloor\!\rceil\  t[i] \neq i  \\
	\ \ \ \longrightarrow \ 
          t[i]^{\oplus\oplus};\ l_F
	\end{array} \right. \\
\end{array}
$
\\	
end loop\ \ \ \ \ \ \ \ \ \ \ \ \ \ \ \ \ \ \ \ \ \ \ \ \ \ \ \ \ \ \ \ \ \ \ \ \ \ \ \ \ \ \ \ \ \ \ \ \ \ \ \ \ \ \ \ \ \ \ \ \ \ \ \ \ \ \ \ \ \ \ \ \ \ \ \ \ \ \ \ \ \ \ \ \ \ \ \ \ \ \ \ \ \ \ \ \ \ \ \\

\caption{The algorithm with $N$ stations and $1$ fault, $M1$.}
\label{pg1}
\end{figure}

The protocol is composed of two inputs and $N$ processes $P[i]$
running in parallel.
The inputs of the protocol are the parameter $N$ and a {\em boolean fault}
which is  {\em true} in case a fault occurs.
Each station $P$
is  a non-deterministic  
finite state
machine extended with 
local variables. The local  variables are the counters $A$ and $F$ for $C\!Acc$
and $C\!F\!ail$, the membership vector $m$,
the variable $t$ to keep track of the time slots, and the
variable $s$ to remember the identity of the station 
which is sending when a fault occurs. Following~\cite{pnueli},
all local variables are marked
with the identity of the station they belong to, which is denoted by  $[i]$.

The  parallel composition is synchronous and 
 the automata  synchronize on
 the broadcast message $emit$.

There are four locations $l_{in}$, $l_0$, $l_1$ and  $l_F$. 
Location $l_{in}$ is the initial location or state. 
When a fault occurs, stations that recognize the fault move
to state  $l_0$ while stations that do not recognize the fault
move to $l_1$. Stations who are prevented from sending
move to location $l_F$.

Let us go through all transitions of location $l_{in}$ in details.
In the initial state  $l_{in}$, all stations have the same
membership vectors. 
First,
suppose that there is no fault.
 A station
can always emit when it is its turn to do so,
expressed in the model by the condition  $t[i] = i$.
The 
first transition has $t[i] = i \ \wedge\  \neg fault$ as guard.
The action performed by that transition is: 
$emit!!, t[i]^{\oplus\oplus}, A[i]:=1;\ l_{in}$.
Put informally, when it is its turn to emit
and there is no fault, a station sends a frame, increments the variable $t$
 by $1$ modulo $N$, resets the counter $A$
 to $1$ and, then, remains in   state  $l_{in}$.
The third transition models the behaviour of a receiving station
when there is no fault.
In that case, it always 
recognizes as correct  frames sent by other stations.
This is expressed in the model by the guard 
$t[i] \neq i \ \wedge\ emit??\ \wedge\  \neg fault$.
These two guards, from the first and third transition,
for two different processes $i$ and $j$ 
are
true simultaneously. Indeed, in the synchronous model of computation,
a computational step 'takes no time'~\cite{ieee}, therefore the
emission of $emit$ by process $i$ is synchronous with its reception by
all other processes. When the parallel statement is finished,
all  stations have incremented $t$ by $1$, the emitting station
has reset $A$ to $1$ while other stations have incremented $A$.
 Counter $F$
stays at its initial value $0$ since it is not incremented in  transitions
containing the condition $\neg fault$.

Suppose now that there is a fault, 
which is  modeled by the condition $ fault$.
For an emitting station, this is the second transition.
The action taken by the emitting station is similar
to the first transition, except that it records its identity
with $s[i]:=t[i]$ and then moves to
location $l_1$ since, by the confidence principle,
a station never thinks of itself as faulty (rather receiving stations
are faulty). 
For a receiving station the guard 
$t[i] \neq i \ \wedge\  emit?? \ \wedge\  fault$ is true.
The occurrence of an asymmetric fault is modeled by a  
non-deterministic choice represented by the fourth and fifth transitions.
The information of whether a station recognizes the fault 
is recorded in the control via  locations $l_0$ 
and $l_1$. If a receiving station recognizes the fault,
 it increases $F$ by $1$,
$ F[i]^{++}$, memorizes the identity of the faulty station, $s[i]:=t$,
puts the membership bit of the sending station to $0$, $m[i][t[i]]:=0$
and moves to $l_0$ (fifth transition).
If a station does not recognize the fault, it increases $A$ by $1$,
$ A[i]^{++}$,
 and moves to $l_1$.

In location  $l_0$, a station behaves as follows.
In its time slot,  $t[i] = i$, either it passes the clique
avoidance mechanism,  $ A[i] > F[i]$ (first transition), and emit, $emit!!$, 
reset its counters,
$A[i]:=1, F[i]:=0$, increments the time slot,   $t[i]^{\oplus\oplus} $,
and stays in $l_0$, or the clique avoidance mechanism fails 
$ A[i] \leq F[i]$ (second transition), 
it cannot emit, $\neg emit!!$, and goes to the
fail state $l_F$.
Outside its time slot, $t[i] \neq i$, if a message
has been broadcast,  $emit??$, and it has the same
membership vector as the sending station,    $m[t[i]] = m[i]$
(third transition),
it increases the counter of accepted messages, $A[i]^{++}$,
increments the time slot,   $t[i]^{\oplus\oplus} $,
and stays in $l_0$; if it does not have the same
membership vector as the sending station,    $m[t[i]] \neq m[i]$ 
(fourth transition),
it increases the counter of failed messages, $F[i]^{++}$,
puts the membership bit of the sending station to $0$, $m[i][t[i]]:=0$,
increments the time slot,   $t[i]^{\oplus\oplus} $,
and stays in $l_0$. Finally, if no message
has been broadcast,  $\neg emit??$ (fifth transition), 
it puts the membership bit of 
the  station that failed sending to $0$, $m[i][t[i]]:=0$,
and stays in $l_0$.

In location  $l_1$, transitions are similar except
that two more transitions are added to cover
the result of  {\tt CheckI} and {\tt CheckII}
using Propositions \ref{leave-active}
and  \ref{leave-active-1}. These are the third and the fourth transitions.
If the sending station is the second successor of the faulty station,
$t[i]=s[i]+2$, and the two successors of the faulty station
have the membership bit of $s[i]$ to $0$, 
$m[t[i]]s[i]=  m[t[i]-1]s[i] = 0$, then station $s[i]$ moves to $l_F$,
otherwise  
station $s[i]$ stays in $l_1$.

In location  $l_F$, stations keep only track of the time slot, they
cannot send and stay there.

\subsection{Construction of an abstract model}

We show in this section the construction of a counter automaton
which is an abstract model of the parametrized membership algorithm 
for an arbitrary number of components $N$. 
In order to simplify the presentation and the proof
of the abstraction, we present this construction in several basic steps.
The aim of the first steps is to encode the infinite-data-domain 
local variables of the $N$ components with a finite number of global variables (counters).
Then, the last step is a counter abstraction which encodes the 
control configurations (for $N$ components) with global variables counting 
the number of components at each control location.

In the sequel, we give the different abstraction steps by defining each time
the abstract model and by showing that it (bi)simulates the previous one.

\subsubsection{Eliminating identical locals $t$ and $s$}

The first transformation we perform is to replace $N$ local
variables $t[i]$ and $s[i]$ by two global counters $t_G$ and $s_G$.
It relies on the following fact:

\begin{lemma}
The two following properties are invariants of the program $M1$:
\begin{enumerate}
\item
$\forall i,j \in \{1 \ldots N\}: (t[i]=t[j]\ \wedge\ s[i]=s[j])$
\item
 $\exists !  i \in \{1 \ldots N\}: t[i]=i$
\end{enumerate}
\end{lemma}

%
%

In other words, at any computation step, all processes $P[i]$ have identical
values for locals $s[i]$ and $t[i]$ and there is exactly 
one process whose identity is $t[i]$.
We define a program $M2$ where these local variables are replaced
by $t_G$ and $s_G$ in the following manner.
We modify the transitions in order to encode the updates
of the local variables as updates of the global ones.
Since all processes are synchronous, these updates must be done
by exactly one component. We choose that this will be done by 
the component for whom it is the turn to emit, i.e., who 
satisfies $t[i]=i$.

So, the program $M2$ is obtained from $M1$ 
by applying the following transformations:
(1) initialize $t_G$ and $s_G$ to 1 and 0 respectively,
(2) replace each occurrence of $t[i]$ in the guards by $t_G$,
and
(3) in all the guarded commands, if $t[i]=i$ appears
in the guard, then replace $t[i]$ by $t_G$ (resp. $s[i]$ by $s_G$),
else remove the update statements of $t[i]$ and $s[i]$.

For example the second transition at location $l_{in}$, see Figure \ref{pg1},
 is transformed into
$$t_G = i \; \wedge \; fault \; \; \longrightarrow \; \; emit!!, t_G^{\oplus\oplus}, 
       A[i]:=1, s_G:=t_G;\ l_1$$

We establish now that  $M2$ bisimulates   $M1$. For that,
let us consider the relation $\alpha_{1,2}$ between states of 
$M1$ and $M2$ such that, for every $\sigma_1$ a state of $M1$, and 
for every $\sigma_2$ a state of $M2$, we have
$\alpha_{1,2}(\sigma_1, \sigma_2)$ if and only if
(1) $\sigma_2$ and $\sigma_1$ coincide on all locations and variables different
from the locals $t$ and $s$, and the globals $t_G$ and $s_G$,
and
(2) $\sigma_2(t_G) = \sigma_1(t[i])$ and $\sigma_2(s_G) = \sigma_1(s[i])$
for every $i \in \{ 1,  \ldots, N \}$, i.e., the value of $t_G$ and $s_G$
in $\sigma_2$ is the same as the value of $t[i]$, respectively $s[i]$
in $\sigma_1$.
Then, it can easily be checked that the relation
$\alpha_{1,2}$ is a bisimulation between the  transition systems
of $M1$ and $M2$. 

\subsubsection{Eliminating locals $A$ and $F$}

Based on Theorem~\ref{countingAcc}, we define a new model
where local variables $A$ and $F$ are simulated by global counters $d_0$, $d_1$, 
$C_0$, and $C_1$;  the counters $C_0$ and $C_1$ stand for $\mid\! S_0\!\mid$ and
$\mid\! S_1\!\mid$ respectively.
We need in addition a variable $r$ to count the current round and 
a variable $C_p$ which counts the number of steps performed in the current round. 

We define hereafter a program $M3$ obtained from $M2$ by the following
transformations:
(1) transitions starting at location $l_{in}$ where the fault is detected ($fault$ appears 
in the guard), replace $A[i]$ and $F[i]$ by $C_1$ and
$C_0$ respectively.
(2) each transition starting from locations $l_0$ and $l_1$
which corresponds to an emit action (where $t_G = i$ appears in the guard)
is duplicated into three transition corresponding  to the cases
where the execution is inside the first round, is inside
some later round, or is precisely at the beginning of a new round.
The comparisons of $A$ and $F$ in the guards and their updates are 
replaced by corresponding comparisons and updates on
$C_0$, $C_1$, $d_0$, and $d_1$, according to Theorem~\ref{countingAcc}.
(3) all other statements involving $A$ and $F$ are removed.

For example the fourth transition at location $l_{in}$~:
$$t_G \neq i \  \wedge\  emit?? \ \wedge\  fault 
          \ \ \  \longrightarrow \
           t_G^{\oplus\oplus}, 
           A[i]:=1;\ l_1$$
    is transformed into
    $$t_G \neq i \  \wedge\  emit?? \ \wedge\  fault 
           \ \ \  \longrightarrow \
               t_G^{\oplus\oplus}, 
          {C_1}^{\oplus\oplus} ;\ l_1.$$
The second transition at  location $l_0$~:\\
$$t_G = i \ \wedge\  A[i] \leq F[i]\ 
 \longrightarrow \ 
      \neg emit!!,  t_G^{\oplus\oplus}; l_F $$
  is duplicated  into three transitions $t_1, t_2, t_3$ as follows:
\begin{center}
   $t_G = i \ \wedge\ C_0+C_1\leq 2\times d_1  \ \wedge\ C_p<N \ \wedge\ r=0\ 
     \longrightarrow \ 
             \neg emit!!,  t_G^{\oplus\oplus}, 
                C_p^{++}, C_0^{--}; l_F $\\
	$t_G = i \ \wedge\  C_0\leq C_1 \ \wedge\ C_p=N \  \longrightarrow \ 
         \neg emit!!,  t_G^{\oplus\oplus}, 
         C_p:=1, r^{++}, C_0^{--}; l_F $.\\
												   $t_G = i \ \wedge\  C_0\leq C_1 \ \wedge\ C_p<N \ \wedge\ r>0\  \longrightarrow \ 
         \neg emit!!,  t_G^{\oplus\oplus}, 
             C_p^{++}, C_0^{--}; l_F $\\
\end{center}

The obtained program $M3$ is bisimilar to $M2$ by Theorem~\ref{countingAcc}.

\subsubsection{Eliminating the local $m$}

At this stage, we can see that the information given by $m$
is not relevant anymore except in the case where
a faulty station must leave the active state because
its first and second successors have recognized it as faulty.
To deal with this case, we replace the test on the vector $m$ by the 
faulty station with a guess of the diagnostic of 
its two immediate successors ({\tt checkI} and {\tt checkII}).
This guess is  made exactly one round after the fault.
For that, we simply perform a nondeterministic choice whether
to leave or to stay in the active state.
It turns out 
that this abstraction
is precise enough for our purpose.

Then, we define a new program $M4$ obtained from the program $M3$
by (1) removing the transitions starting from $l_1$ involving tests on 
the membership bit vectors of the two immediate successors,
(2) duplicating the transitions from  $l_1$ for the case $(C_p = N) \wedge (r = 0)$ 
into two transitions corresponding to the actions of leaving or staying 
in the active state with $g$ to fix the choice, and (3) removing all the remaining statements involving $m$.

Moreover, after the transformation above, guarded commands starting at
$l_0$ and $l_1$ which involve the test $t_G \neq i$ can actually be compacted
into trivial self loops. This leads to the program $M4$ 
given in Figure~\ref{pg6}. 
One shows that this program simulates the program $M3$ by induction.

\begin{figure}
$
\begin{array}{lcllcllcllcl}
$in $N & :& $int, $N > 1 & \    C_0 & : &  $int, init $0 & 
\   t_G & : & $int, init $1  & \    r& : & $int, init $0  \\
$in $fault & : & $bool. signal$ & \  C_1 &: & $int, init $1 &
\  s_G & : &  $int, init $0 & \   d_0 & : &  $int, init $0 \\
$in $g & :& $bool. signal$ &\   C_p & : & $int, init $1 &
\  d_1 & : & $int, init $1 &\ \ \\
\end{array}
$

\medskip

$\parallel_{i=1}^N P[i]$, \ \ \ 
where $P[i]$ is the following program:


loop forever do  \ \ \ \ \ \ \ \ \ \ \ \ \ \ \ \ \ \ \ \ \ \ \ \ \ \ \ \ \ \ \ \ \ \ \ \ \ \ \ \ \ \ \ \ \ \ \ \ \ \ \ \ \ \ \ \ \ \ \ \ \ \  \\
$
\begin{array}{ll}
 \ \   l_{in} & \left \{ \begin{array}{l}
                    \lfloor\!\rceil\ t_G = i \ \wedge\  \neg fault \\
	\ \ \ 	    \longrightarrow \ 
         emit!!, t_G^{\oplus\oplus};\ l_{in} \\
	 \lfloor\!\rceil\ t_G = i \ \wedge\  fault \\   
	 \ \ \ \longrightarrow \ 
	 emit!!, t_G^{\oplus\oplus}, s_G:=t_G;\ l_1 \\
	 \lfloor\!\rceil\ t_G \neq i \ \wedge\  emit??\ \wedge\  \neg fault \\  
	 \ \ \ \longrightarrow\ 
           ;\ l_{in} \\
	  \lfloor\!\rceil\ t_G \neq i \ \wedge\  emit?? \ \wedge\  fault\\  
	  \ \ \  \longrightarrow \   C_1^{++};\ l_1 \\
	  \lfloor\!\rceil\ t_G \neq i \ \wedge\  emit?? \ \wedge\  fault  
	  \ \ \ \longrightarrow \   C_0^{++};\ l_0 \\
	  \end{array} \right. \\
\\
\ \     l_0 & \left \{ \begin{array}{l}
    \lfloor\!\rceil\ t_G = i \ \wedge\ C_0+C_1>2\times d_1 \ \wedge\ C_p<N \ \wedge\ r=0\\ 
     \ \ \ \longrightarrow \ 
         emit!!,  t_G^{\oplus\oplus}, 
	  C_p^{++}, d_0^{++};\ l_0 \\
     \lfloor\!\rceil\ t_G = i \ \wedge\  C_0>C_1 \ \wedge\ C_p=N  \\ 
     \ \ \ \longrightarrow \ 
         emit!!,  t_G^{\oplus\oplus}, 
	  C_p:=1, r^{++};\ l_0 \\
     \lfloor\!\rceil\ t_G = i \ \wedge\  C_0>C_1 \ \wedge\ C_p<N \ \wedge\ r>0\\  
     \ \ \ \longrightarrow \ 
         emit!!,  t_G^{\oplus\oplus}, 
	  C_p^{++};\ l_0 \\
	 \lfloor\!\rceil\ t_G = i \ \wedge\  C_0+C_1\leq 2\times d_1 \ \wedge\ C_p<N \  
	 \wedge\ r=0\ \\ 
	 \ \ \ \longrightarrow \ 
          \neg emit!!,  t_G^{\oplus\oplus}, C_p^{++}, C_0^{--};\ l_F \\
    \lfloor\!\rceil\ t_G = i \ \wedge\ C_0\leq C_1 \ \wedge\ C_p=N \\ 
    \ \ \ \longrightarrow \ 
          \neg emit!!,  t_G^{\oplus\oplus}, C_p:=1, r^{++}, C_0^{--};\ l_F \\
    \lfloor\!\rceil\ t_G = i \ \wedge\ C_0\leq C_1 \ \wedge\ C_p<N \ \wedge\ r>0\\   
    \ \ \ \longrightarrow \ 
          \neg emit!!,  t_G^{\oplus\oplus}, C_p^{++}, C_0^{--};\ l_F \\
	  \lfloor\!\rceil\ t_G \neq i \ 
      \ \ \ \longrightarrow \ ;\ l_0
	\end{array} \right. \\
\\
\ \    l_1 & \left \{ \begin{array}{l}
     \lfloor\!\rceil\ t_G = i \ \wedge\ C_0+C_1> 2\times d_0 \ \wedge\ C_p<N \wedge r=0\\  
\ \ \ \longrightarrow \ 
         emit!!,  t_G^{\oplus\oplus}, 
	 C_p^{++}, d_1^{++};\ l_1 \\
     \lfloor\!\rceil\ t_G = i\ \wedge\ C_1 > C_0\ \wedge\ C_p=N\ \wedge 
	r \neq 0  \\   
     \ \ \ \longrightarrow \ 
         emit!!,  t_G^{\oplus\oplus}, 
	 C_p:=1, r^{++};\ l_1 \\
\lfloor\!\rceil\ t_G = i \ \wedge\ C_1 > C_0  \ \wedge\ C_p=N\ 
	\wedge r=0\ \wedge \ \neg g\\  
     \ \ \ \longrightarrow \ 
 emit!!,  t_G^{\oplus\oplus}, 
	 C_p:=1, r^{++};\ l_1 \\
    \lfloor\!\rceil\ t_G = i \ \wedge\ C_1 > C_0  \ \wedge\ C_p=N\ 
	\wedge r=0\ \wedge \  g\\  
     \ \ \ \longrightarrow \ 
       \neg emit!!,  t_G^{\oplus\oplus}, 
	 C_p:=1, r^{++},  C_1^{--};\ l_F \\
    \lfloor\!\rceil\ t_G = i \ \wedge\ C_1 > C_0  \ \wedge\ C_p<N \ \wedge\ r>0\\   
     \ \ \ \longrightarrow \ 
         emit!!,  t_G^{\oplus\oplus}, 
	 C_p^{++};\ l_1 \\
  \lfloor\!\rceil\ t_G = i \ \wedge\ C_0+C_1\leq 2\times d_0  \ \wedge\ C_p<N \wedge r=0\\ 
	\ \ \  \longrightarrow \ 
          \neg emit!!,  t_G^{\oplus\oplus}, C_p^{++}, C_1^{--};\ l_F \\
	   \lfloor\!\rceil\ t_G = i \ \wedge\  C_1 \leq C_0 \ \wedge\ C_p=N \\
	\ \ \  \longrightarrow \ 
          \neg emit!!,  t_G^{\oplus\oplus},C_p:=1, r^{++}, C_1^{--};\ l_F \\
	\lfloor\!\rceil\ t_G = i \ \wedge\  C_1 \leq C_0 \ \wedge\ C_p<N\ \wedge\ r>0\\   
	\ \ \  \longrightarrow \ 
          \neg emit!!,  t_G^{\oplus\oplus},C_p^{++}, C_1^{--};\ l_F \\
	  \lfloor\!\rceil\ t_G \neq i \
	  \ \ \   \longrightarrow \ ;\ l_1
	\end{array} \right. \\
\\
\ \    l_F & \left \{ \begin{array}{l}
       \lfloor\!\rceil\ t_G = i \ \wedge\ C_p<N \\  
       \ \ \ \longrightarrow \ 
         \neg emit!!, t_G^{\oplus\oplus}, C_p^{++};\ l_F\\
      \lfloor\!\rceil\ t_G = i \ \wedge\ C_p=N \\   
      \ \ \ \longrightarrow \ 
         \neg emit!!, t_G^{\oplus\oplus}, C'_p:=1, r^{++};\ l_F\\
	\lfloor\!\rceil\   t_G \neq i \
	\ \ \ \longrightarrow \ 
          ;\ l_F
	\end{array} \right. \\
\end{array}
$
\\
end loop\ \ \ \ \ \ \ \ \ \ \ \ \ \ \ \ \ \ \ \ \ \ \ \ \ \ \ \ \ \ \ \ \ \ \ \ \ \ \ \ \ \ \ \ \ \ \ \ \ \ \ \ \ \ \ \ \ \ \ \ \ \ \ \ \ \ \ \ \  \\
\caption{Eliminating all locals, $M4$.}
\label{pg6}
\end{figure}

\subsubsection{Strengthening guards}

Before the counter abstraction step where identities of processes
will be lost, we need to strengthen the guards using some invariants of the system.

\begin{lemma}
\label{invariants}
The following statements are invariants of the program $M4$:
\begin{enumerate}
\item 
$\forall i \;
 (l_0[i]\ \wedge\ t_G=i\  \wedge\ C_0+C_1>2\times d_1 
 \ \wedge\ C_p<N \ \wedge\ r=0 )\ \Rightarrow \ d_0 < C_0\ )$
\item 
 $\forall i \;
(l_0[i]\ \wedge \ t_G = i \ \wedge\  C_0+C_1\leq 2\times d_1 \ \wedge\ C_p<N \   \wedge\ r=0)\ \Rightarrow \ d_0 < C_0\ )$
\item
 $\forall i \;
(l_1[i]\ \wedge \ t_G = i \ \wedge\ C_0+C_1> 2\times d_0 \ \wedge\ C_p<N \wedge \ r=0)\ \Rightarrow \ d_1 < C_1\ )$
\item
 $\forall i \;
(l_1[i]\ \wedge \ t_G = i \ \wedge\ C_0+C_1\leq 2\times d_0  \ \wedge\ C_p<N \wedge\ r=0)\ \Rightarrow \ d_1 < C_1\ )$ 
\end{enumerate}
\end{lemma}

Invariant (1) in the lemma above says that when process $P[i]$ is at location $l_0$,
if it is the turn of this process to emit, 
and if it is allowed to emit in the round following the fault, then 
not all stations from the set $S_0$ have emitted in that round.
This is due to the fact that $d_0$ counts stations from the set $S_0$ 
that have emitted in that current round. 

Thus, at location $l_0$, the 
guard $t_G=i\  \wedge\ C_0+C_1>2\times d_1 \ \wedge\ C_p<N \ \wedge\ r=0$ can be strengthened 
into $t_G=i\  \wedge\ C_0+C_1>2\times d_1 \ \wedge\ C_p<N \ \wedge\ r=0 \ \wedge\ d_0 < C_0$ 
without changing the semantics of the program. We do similar transformation
using the other invariants.

Further, we update $d_0$ and $d_1$ in other
transitions at $l_0$ and $l_1$ so that they keep counting the number
of  stations from the set $S_0$, $S_1$ respectively,
that have emitted in subsequent rounds. In that way, we obtain more
invariants similar to the ones given in Lemma~\ref{invariants}
and we use them to strengthen guards without 
changing the semantics of the program.

For example, at $l_0$ transition
$$t_G = i \ \wedge\  C_0>C_1 \ \wedge\ C_p=N   
     \ \longrightarrow \ 
         emit!!,  t_G^{\oplus\oplus},  
	  C_p:=1, r^{++};\ l_0$$

is changed into
$$t_G = i \ \wedge\  C_0>C_1 \ \wedge\ C_p=N   
     \ \longrightarrow \ 
         emit!!,  t_G^{\oplus\oplus}, d_0:=0, 
	  C_p:=1, r^{++};\ l_0$$
and transition
$$t_G = i \ \wedge\  C_0>C_1 \ \wedge\ C_p<N \ \wedge\ r>0\\  
     \ \ \ \longrightarrow \ 
         emit!!,  t_G^{\oplus\oplus},
	  C_p^{++};\ l_0$$

is changed into
$$t_G = i \ \wedge\  C_0>C_1 \ \wedge\ C_p<N \ \wedge\ r>0\\  
     \ \ \ \longrightarrow \ 
         emit!!,  t_G^{\oplus\oplus},d_0^{++}, 
	  C_p^{++};\ l_0.$$

It can be shown that
$$\forall i \;
 (l_0[i]\ \wedge\ t_G = i \ \wedge\  C_0>C_1 \ \wedge\ C_p=N ) 
\ \Rightarrow \ d_0 = C_0\ )$$
is an invariant and therefore the guard can be strengthened into
$$t_G = i \ \wedge\  C_0>C_1 \ \wedge\ C_p=N \ \wedge \ d_0 = C_0.$$

Similarly to transitions starting from $l_0$ and $l_1$
we need to strengthen the guards of the transitions starting from 
$l_F$.  For that, we introduce two supplementary counters $d_F$ and $C_F$ --initial value 0--
which play a similar r\^ole for location $l_F$ as counters $d_0$ and  $C_0$
do for location $l_0$.
We use  invariants similar to those given in the lemma above to strengthen all guards at $l_0$, $l_1$ and $l_F$.

The last guard strengthening we perform makes use of  
Proposition \ref{leave-active} that says
that if the faulty station leaves the active state because
of its first and second successor, then the two stations
following it do not belong to $S_1$, but to $S_0$.
In other words, a station from $S_1$ emitting in the first round following the fault with $g$ cannot be the first nor the second station following
the faulty station. 
Therefore, at location $l_1$,  the guard of transition 
$$\ t_G = i \ \wedge\ C_1 > C_0  \ \wedge\ C_p<N \ \wedge\ r>0\\   
     \  \longrightarrow \ 
         emit!!,  t_G^{\oplus\oplus}, 
	 C_p^{++};\ l_1$$
is strengthened into

$t_G = i \ \wedge\ C_1 > C_0  \ \wedge\ C_p<N \ \wedge\ r>0 \wedge
(r\neq 1 \vee \neg g \vee C_p\neq 1\vee C_p\neq 2)$ without changing
the semantics of the program.

We obtain a transformed
program $M5$ bisimilar to program $M4$. 

\subsubsection{Counter abstraction}

 We are now ready to  perform the usual counter abstraction 
where individual
control locations $l_{in}$, $l_0$, $l_1$ and $l_F$
 are replaced by counters $C_{in}$, $C_0$, $C_1$ and $C_F$ respectively
counting 
how many processes 
are  at these locations and obtain a single extended finite state machine
with a single location, which is then omitted~\cite{delzano}.

Then, we define a new program $M6$ obtained from the program $M5$
by (1) transforming 
transition $l : g \ \rightarrow \ a; l'$  into
$C_l >0 \ \wedge\  g \ \rightarrow \ a, C_l^{--}, C_{l'}^{++}$,
where $l, l'$ are locations or control states and $C_l$,$C_{l'}$ 
the associated counters, (2) replacing  
any condition involving $i$  by {\em true}, (3) packing into
a single abstract transition
 all  transitions from $M5$ where guards evaluate
to true simultaneously.

For example, the two  transitions of $M5$ at  $l_{in}$
 $$t_G = i \ \wedge\  \neg fault \ \longrightarrow \ 
         emit!!, t_G^{\oplus\oplus}; l_{in} $$
and
$$t_G \neq i \ \wedge\  emit??\ \wedge\  \neg fault \  
	 \longrightarrow\ 
           ; l_{in} $$  
give the abstract transition: 
$$C_{in} >0 \ \wedge\ emit??\ \wedge\  \neg fault 
		    \ \longrightarrow \ emit!!, 
          t_G^{\oplus\oplus}.$$
We obtain  
program $M6$ shown in Figure~\ref{pg8}.
The initial value of $x$ guesses
the number of processes  that move from location
$l_{in}$ to $l_1$.

Such a counter abstraction simulates the previous models and
verifies the property that the value
of any counter $C_l$ at some state $\sigma'$ is exactly the number
of processes at location $l$ in the corresponding state $\sigma$.

 Thus, if we show that $C_1$
or $C_0$ evaluate to $0$ in the program given in Figure~\ref{pg8}, 
we know that set
$S_0$ or $S_1$ is empty in the concrete model, establishing that there
is no clique.

\begin{figure}
$
\begin{array}{lcllcllcllcl}
$in $N & :& $int, $N > 1 &\   C_0& : & $int, init $0 & 
\  t_G& :&  $int, init $1 &\  r& :&  $int, init $0\\
$in $fault& :& $bool. signal$ &\   C_1& :&  $int, init $1 &
\    s_G& :&  $int, init $0 &\   d_0& :&  $int, init $0\\
$in $g& :& $bool. signal$ &\   C_p& :&  $int, init $1 &
\  d_1& :&  $int, init $1 &\   d_F& :&  $int, init $0 \\
$in $x$$& :& $int, $x \geq 1 &\   C_i& :&  $int, init $N & 
\  C_F& : & $int, init $0 &\  \\
\end{array}
$

loop forever do \ \ \ \ \ \ \ \ \ \ \ \ \ \ \ \ \ \ \ \ \ \ \ \ \ \ \ \ \ \ \ \ \ \ \ \ \ \ \ \ \ \ \ \ \ \ \ \ \ \ \ \ \ \ \ \ \ \ \ \ \ \ \ \ \ \ \ \ \ \ \ \ \ \ \ \ \ \ \ \ \ \ \ \ \ \ \ \ \ \ \ \ \  \\
$
\begin{array}{ll}
 \   & \left \{ \begin{array}{l}
                    \lfloor\!\rceil\ C_i >0 \ \wedge\ emit??\ \wedge\  \neg fault \\
		    \ \ \ \longrightarrow \ emit!!, 
          t_G^{\oplus\oplus} \\
	 \lfloor\!\rceil\ C_i >0 \ \wedge\ emit??\ \wedge\ fault \\  
	\ \ \ \ \longrightarrow \ 
	  emit!!, t_G^{\oplus\oplus}, s_G':=t_G, C'_1:= x, 
	  C'_0:= N-x, C'_i:= 0 \\
 \lfloor\!\rceil\ C_0 >0 \ \wedge\ d_0<C_0 \ \wedge\ C_0+C_1>2\times d_1 \ 
\wedge\ C_p<N \ \wedge\ r=0\\ 
     \ \ \ \longrightarrow \ 
         emit!!,  t_G^{\oplus\oplus}, 
	  C_p^{++}, d_0^{++} \\
     \lfloor\!\rceil\ C_0 >0 \ \wedge\ d_0=C_0 \ \wedge\  C_0>C_1 \ \wedge\ C_p=N  \\ 
     \ \ \ \longrightarrow \ 
         emit!!,  t_G^{\oplus\oplus}, 
	  C'_p:=1, r^{++}, d'_0:=1, d'_1:=0, d'_F:=0  \\
     \lfloor\!\rceil\ C_0 >0 \ \wedge\ d_0<C_0 \ \wedge\  C_0>C_1 \ \wedge\ C_p<N 
\ \wedge\ r>0\\  
     \ \ \ \longrightarrow \ 
         emit!!,  t_G^{\oplus\oplus}, 
	  C_p^{++}, d_0^{++}  \\
	 \lfloor\!\rceil\ C_0 >0 \ \wedge\ d_0<C_0 \ \wedge\  C_0+C_1\leq 2\times d_1 
	 \ \wedge\ C_p<N \  
	 \wedge\ r=0\ \\ 
	 \ \ \ \longrightarrow \ 
      \neg emit!!,  t_G^{\oplus\oplus}, C_p^{++}, C_0^{--}, C_F^{++}, d_F^{++} \\
    \lfloor\!\rceil\ C_0 >0 \ \wedge\ d_0=C_0\ \wedge\ C_0\leq C_1 \ \wedge\ C_p=N \\ 
    \ \ \ \longrightarrow \   \ 
 \neg emit!!,  t_G^{\oplus\oplus}, C'_p:=1, r^{++}, C_0^{--}, C_F^{++}, 
 d'_0:=0, d'_F:=1, d'_1:=0 \\
    \lfloor\!\rceil\ C_0 >0 \ \wedge\ d_0<C_0 \ \wedge\ C_0\leq C_1 
    \ \wedge\ C_p<N \ \wedge\ r>0\\   
    \ \ \ \longrightarrow \ 
    \neg emit!!,  t_G^{\oplus\oplus}, C_p^{++}, C_0^{--}, C_F^{++}, d_F^{++} \\
    \lfloor\!\rceil\ C_1 >0 \ \wedge\ d_1<C_1\ \wedge\ C_0+C_1> 2\times d_0 
\ \wedge\ C_p<N \wedge r=0\\  
\ \ \ \longrightarrow \ 
         emit!!,  t_G^{\oplus\oplus}, 
	 C_p^{++}, d_1^{++}\\
    \lfloor\!\rceil\ C_1 >0\ \wedge\ d_1=C_1\ \wedge\ C_1 > C_0\ \wedge\ C_p=N\ 
	\wedge (r\neq 0 \vee \neg g)\\   
   \ \ \   \longrightarrow \ 
         emit!!,  t_G^{\oplus\oplus}, 
	 C'_p:=1, r^{++}, d'_1:=1, d'_0:=0, d'_F:=0  \\
     \lfloor\!\rceil\ C_1 >0\ \wedge\ d_1=C_1\ \wedge\ C_1 > C_0  \ \wedge\ C_p=N \ 
	\wedge r=0 \wedge g\\  
     \ \ \ \longrightarrow \ 
         \neg emit!!,  t_G^{\oplus\oplus}, 
	 C'_p:=1, r^{++}, d'_1:=0, d'_F:=1, d'_0:=0, C_F^{++}  \\
    \lfloor\!\rceil\  C_1 >0\ \wedge\ d_1<C_1 \ \wedge\ C_1 > C_0  
    \ \wedge\ C_p<N \ \wedge\ r>0\ \wedge \\
   \ \ \ \ 
	(r\neq 1 \vee \neg g \vee C_p\neq 1\vee C_p\neq 2) \\  
     \ \ \ \longrightarrow \  
         emit!!,  t_G^{\oplus\oplus}, 
	 C_p^{++}, d_1^{++}  \\
  \lfloor\!\rceil\  C_1 >0\  \wedge\ d_1<C_1\ \wedge\ C_0+C_1\leq 2\times d_0  
  \ \wedge\ C_p<N \wedge r=0\\ 
\ \ \ 	 \longrightarrow \ 
      \neg emit!!,  t_G^{\oplus\oplus}, C_p^{++}, C_1^{--}, C_F^{++}, d_F^{++}\\
      \lfloor\!\rceil\  C_1 >0 \ \wedge\ d_1=C_1\ \wedge\  C_1 \leq C_0 \ \wedge\ C_p=N \\
	 \ \ \ \longrightarrow \ 
  \neg emit!!,  t_G^{\oplus\oplus},C'_p:=1, r^{++}, C_1^{--}, C_F^{++}, 
  d'_F:=1, d'_1:=0, d'_0:=0\\
	   \lfloor\!\rceil\ C_1 >0 \  \wedge\ d_1<C_1\ \wedge\  C_1 \leq C_0 
	   \ \wedge\ C_p<N\ \wedge\ r>0\\ 
	 \ \ \ \longrightarrow \ 
      \neg emit!!,  t_G^{\oplus\oplus},C_p^{++}, C_1^{--}, C_F^{++}, d_F^{++}\\
      \lfloor\!\rceil\ C_F >0\ \wedge\ d_F<C_F \ \wedge\ C_p<N \\  
      \ \ \ \longrightarrow \ 
         \neg emit!!, t_G^{\oplus\oplus}, C_p^{++}, d_F^{++}\\
       \lfloor\!\rceil\  C_F >0 \ \wedge\ d_F=C_F \ \wedge\ C_p=N \\  
       \ \ \ \longrightarrow \ 
         \neg emit!!, t_G^{\oplus\oplus}, C'_p:=1, r^{++}, d'_F:=1, d'_0:=0,
	d'_1:=0 
	\end{array} \right. \\
\end{array}
$
\\	
end loop\ \ \ \ \ \ \ \ \ \ \ \ \ \ \ \ \ \ \ \ \ \ \ \ \ \ \ \ \ \ \ \ \ \ \ \ \ \ \ \ \ \ \ \ \ \ \ \ \ \ \ \ \ \ \ \ \ \ \ \ \ \ \ \ \ \ \ \ \ \ \ \ \ \ \ \ \ \ \ \ \ \ \ \ \ \ \ \ \ \ \ \ \ \ \ \ \ \\
\caption{The  abstract model $M6$.}
\label{pg8}
\end{figure}

\subsection{Proving properties}

We have used 
 the system given Figure~\ref{pg8}, with a slight change concerning
rounds,
 to prove automatically properties
of the protocol. We have used an automaton
where rounds are represented by  states, rather
than by the variable $r$.  We make  the distinction
between  the first round after the fault, and later rounds.
Condition $r=0$ is equivalent to the automaton
being in state {\tt round1} while $r>0$ is equivalent to the automaton
being in state {\tt later round}. Further, the 
automaton leaves state {\tt later round} 
to go to a state {\tt normal} when cliques are stable.

A first property, called $M1$, that has been proved as true, is 
that at the end of the first round
after the fault:
 $$  !(C_1 = C_0)\ \ \ \ \ \ \ \ \ \ (P1).$$

P1 means that, when the first round after the fault is over, 
either $\mid\! S_1\!\mid >\mid\! S_0\!\mid$ 
or $\mid\!S_0\!\mid > \mid\! S_1\!\mid$, whatever the original partition $\{S_1, S_0\}$ was when the fault
occurred. 

We have analyzed what leads to $C_1 > C_0$, or $C_0 > C_1$ 
after $1$ round.

First,  we have shown that, if $\mid\! S_1\!\mid >\mid\! S_0\!\mid$
when the fault occurs, then $C_1 > C_0$ after one round, and vice-versa. 
 Adding the constraint $x > N-x$, we have proved that, at the end
of the first round after the fault :

$$ (x=C_1)\ \ \ \ \ \ \ \ \ \  \ \  (P2).$$

Since counters $C_1$ and $C_0$ may not increase,  this implies
$C_1 > C_0$ when $r>0$.
It also implies that all stations from $S_1$ did send in the
first round.

Then we have investigated the  case 
$\mid\! S_1\!\mid = \mid \! S_0\!\mid$ 
   when the fault occurs.
If set $S_1$ comes first in the statical order,  
then $C_1 >C_0$, and vice versa
if $S_0$ comes first. Adding the constraint
$x = N-x$ we have proved:

$$AG\ \ ((d_1=x\ {\tt and}\ d_0<x) \Rightarrow AG\ (C_1=x))\ \ \ \ \ \ (P3),$$

$$ AG\ \ ((d_1=x\ {\tt and}\ d_0 < x)  \Rightarrow  (C_1+C_0-2*d_1<=0))
\ \ \ \ \ \ (P4).$$

Again, this implies
$C_1 > C_0$ after the first round.
It also implies that all stations from $S_1$ did send in the
first round.

To prove the main property, we have first shown that,
if $r>0$, when one round is completed, it is not
possible to start a new round where both $C_1$ and $C_0$
are not $0$, i.e.,  $S_1$ and $S_0$ are both not empty.

Indeed the property below is true when $r>0$:

$$AG \  \neg(\neg (C_1=0) \ {\tt and}\ \neg (C_0=0) \ {\tt and}\ (C_p=N))
\ \ \ \ \ \ \ \ \ \ (P6).$$

Finally, we proved that, at the end of the second round
after the fault, i.e., when $C_p = N$ and the
automaton goes to state {\tt normal}:  

$$AG \  (C_1=0 \ {\tt or}\  C_0=0)\ \ \ \ \ \ \ \ \ \ (P7).$$

$P7$ means   that either  $S_1$
or  $S_0$ is empty at the end of the second round. 
Hence, all active stations have the same membership vectors at the end
of the second round and form again a single clique in the graph
theoretical sense.

\section{Automatic Verification: the $k$ Faults Case}
\label{autk}
To be able to calculate precisely $C\!Acc_s$ and $C\!F\!ail_{s}$ using global counters only, see Theorem \ref{countingAcc}, is what makes possible the construction of an abstract model in the $1$-fault case. For the $k$-faults case, the same approach can be taken. 
Provided that one is able to calculate precisely $C\!Acc_s$ and $C\!F\!ail_{s}$
using global counters, first a behavioural model for a scenario of the $k$-fault case is constructed, then this behavioural model can be transformed into successive models that simulate each other, replacing local variables with global counters, till the final counter abstraction is obtained and verified automatically, as has been done for the $1$-fault case.
Thus, what is needed is to establish a generalisation of Theorem \ref{countingAcc} for the $k$-faults case. 

\subsection{Calculating $C\!Acc_s$ and $C\!F\!ail_{s}$}

Let $ 1 \leq i \leq k$. 
By  Proposition \ref{justafters_kInt}, after the occurrence
of fault $i$, $W$, the set of active stations, is partitioned into sets $S_w$
with $w \in \{0, 1 \}^i$. We find it handy for the following
to indicate the length of the string $w$ with
the superscript $i$. We associate two counters $Cw^i$ and 
$dw^i$ to each set $S_{w^i}$ that is formed after the occurrence of any 
fault $i$. 
The counters $Cw^i$ counts how many stations belong
to set $S_{w^i}$ when fault
$i$ occurs. 
The counters
$dw^i$ count how many stations from the set $S_{w^i}$
have sent between fault $i$ and fault $i+1$  
in case $i < k$, and counts how many stations from the set $S_{w^i}$
have sent so far in the first round following fault $k$ in case $i = k$.
Again  because of  Proposition \ref{justafters_kInt},
we assume  that, for any $ w \in \{0, 1 \}^{i-1}$,
  $Cw^{i-1}1 + Cw^{i-1}0 = Cw^{i}$,  $Cw^{i-1}1 \geq 1$ and $dw^{i-1}1 \geq 1$.
Further, for each fault $i$, we associate a counter $Cp(i)$ that counts
how many time slots have elapsed since fault $i$.

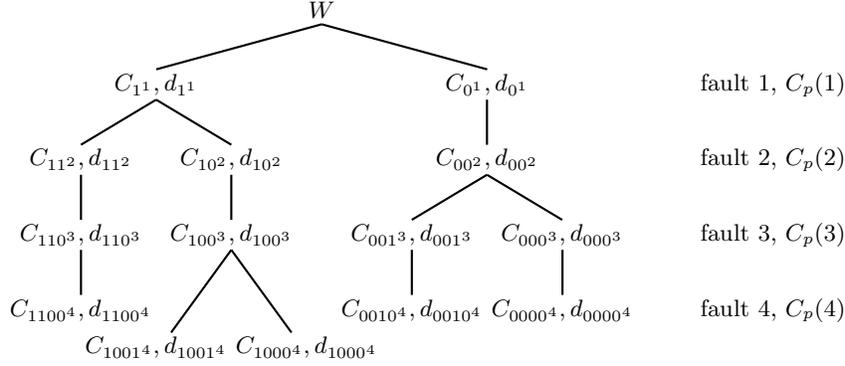
\begin{figure}
\begin{pspicture}(10,5)

\rput(4,5){$W$}

\rput(10,4){fault 1, $C_p(1)$}

\rput(1.8,4){$C_{{1}^1}, d_{{1}^1}$}
\rput(6.2,4){$C_{{0}^1}, d_{{0}^1}$}

\rput(4,4.8){\psline(-2.2,-0.6)}
\rput(4,4.8){\psline(2.2,-0.6)}

\rput(10,3){fault 2, $C_p(2)$}

\rput(6.2,3){$C_{{00}^2}, d_{{00}^2}$}

\rput(6.2,3.8){\psline(0,-0.6)}

\rput(0.8,3){$C_{{11}^2}, d_{{11}^2}$}
\rput(2.8,3){$C_{{10}^2}, d_{{10}^2}$}

\rput(1.8,3.8){\psline(-1,-0.6)}
\rput(1.8,3.8){\psline(1,-0.6)}

\rput(10,2){fault 3, $C_p(3)$}

\rput(0.8,2){$C_{{110}^3}, d_{{110}^3}$}
\rput(2.8,2){$C_{{100}^3}, d_{{100}^3}$}

\rput(0.8,2.8){\psline(0,-0.6)}
\rput(2.8,2.8){\psline(0,-0.6)}

\rput(5.2,2){$C_{{001}^3}, d_{{001}^3}$}
\rput(7.2,2){$C_{{000}^3}, d_{{000}^3}$}

\rput(6.2,2.8){\psline(-1,-0.6)}
\rput(6.2,2.8){\psline(1,-0.6)}

\rput(10,1){fault 4, $C_p(4)$}

\rput(0.8,1){$C_{{1100}^4}, d_{{1100}^4}$}
\rput(0.8,1.8){\psline(0,-0.6)}

\rput(5.2,1){$C_{{0010}^4}, d_{{0010}^4}$}
\rput(7.2,1){$C_{{0000}^4}, d_{{0000}^4}$}

\rput(5.2,1.8){\psline(0,-0.6)}
\rput(7.2,1.8){\psline(0,-0.6)}

\rput(1.8,0.5){$C_{{1001}^4}, d_{{1001}^4}$}
\rput(3.8,0.5){$C_{{1000}^4}, d_{{1000}^4}$}
\rput(2.8,1.8){\psline(-0.8,-1.1)}
\rput(2.8,1.8){\psline(0.8,-1.1)}

\end{pspicture}
\caption{Illustrating the creation of counters up to 4 faults.}
\label{countersto4}
\end{figure}

\begin{figure}
\centering\epsfig{file=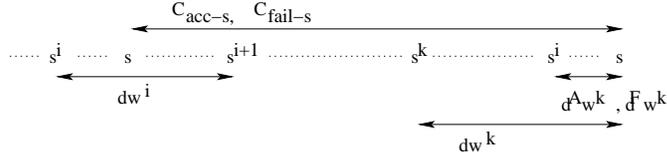, height=2cm}
\caption{Evaluating $C\!Acc_s$ and $C\!F\!ail_{s}$ after fault $k$.}
\label{countAccFork}
\end{figure}

The creation of counters is illustrated taking a particular scenario in Figure \ref{countAccFork}.
After 1 fault, active stations split into two sets, $S_0$, the stations that have not received the frame correctly and $S_1$ the stations that have  received the frame correctly. Four counters are created: $C_{0^1}$ contains the number of stations from $S_0$,  $C_{1^1}$ contains the number of stations from $S_1$,  $d_{0^1}$ counts the stations from $S_0$ that are sending frames, $d_{1^1}$ counts the stations from $S_1$ that are sending frames. $d_{1^1}\geq 1$ since the station that was sending when the fault occurred is from $S_1$. 
Each time a station from $S_1$, respectively from $S_0$ is prevented from sending, then $S_1$ and $C_{1^1}$, respectively $S_0$ and $C_{0^1}$, are decreased by $1$. Suppose that a second fault occurs when some station from $S_1$ emits. Then $S_1$ splits into $S_{10}$ and $S_{11}$ and  six new counters are created: $C_{{00}^2}$, which is, in this scenario, initially equal to the present value of $C_{0^1}$,
$C_{{10}^2}$, to count the number of stations in $S_{10}$, $C_{{11}^2}$ to count
the number of stations in $S_{11}$, $d_{{00}^2}$ to count the number of stations from $S_{00}$ (which is, in this scenario, the same as  $S_{0}$) that are sending frames after the second fault -- note that $d_{0^1}$ is not incremented anymore after the occurence of fault 2-- $d_{{10}^2}$ to count the number of stations from $S_{10}$ that are sending frames after the second fault, and $d_{{11}^2}$ to count the number of stations from $S_{11}$ that are sending frames after the second fault.
Again, $d_{{11}^2}\geq 1$ since the station that was sending when the fault occured is from  $S_{11}$. In Figure \ref{countersto4}, one assumes that a third  fault occurs when a station from set $S_{00}$ is sending and a fourth fault occurs when a station from set $S_{100}$ is sending.

Thus, for $k$ faults and a particular scenario, $\Sigma_{i=1}^{i=k} 2(i+1) \ + k$ counters so far are created.

These counters are almost enough to know  
$C\!Acc_s$ and $C\!F\!ail_{s}$ for any station $s$ in the 
first round following
fault $k$. Indeed, let $s$ be a station ready to send.
$s$ belongs to some set $S_{w^k}$. 
In the rounds preceding fault $k$ and during
the round following fault $k$, $s$ recognizes
as correct  frames sent by stations from $S_{w'}$, where $w'$ is a prefix
of $w^k$, and recognizes as incorrect  all other frames.
This information is recorded with the counters $dw^i$, $w \in  \{0, 1 \}^i$
and $ 1 \leq i \leq k$.

There is one more subtlety. 
The clique avoidance mechanism needs
that  $C\!Acc_s$ and $C\!F\!ail_{s}$ count one round only,
the round being relative
to the position of the sending station $s$. To do so properly,
we distinguish two  cases.

The first case is when fault $k$ occurs in the first round following fault
$1$ and the time slot of $s$ still belongs 
to that round. 
One must take into account that $s$ has recognized as correct all
stations that have sent before fault $1$, which is a straight generalization of
 Theorem \ref{countingAcc}. For example, consider the scenario of Figure \ref{countersto4} and a station s from set $S_{0010}$.  
Then:\\
 $C\!Acc_s = \mid\! W\!\mid - d_{{1}^1} - d_{{11}^2} - d_{{10}^2} 
- d_{{110}^3} - d_{{100}^3} - d_{{000}^3} - d_{{1100}^4} - 
d_{{1001}^4} - d_{{1000}^4} - d_{{0000}^4}$,\\
$C\!F\!ail_{s} = d_{{1}^1} + d_{{11}^2} + d_{{10}^2} 
+ d_{{110}^3} + d_{{100}^3} + d_{{000}^3} + d_{{1100}^4} +
d_{{1001}^4} + d_{{1000}^4} + d_{{0000}^4}$.

The second case is when the time slot of the sending station $s$ lies
between station $s^i$ and $s^{i+1}$ and fault $k$ occurs in the first
 round following fault $i$, $i> 1$. After fault $i$, $s$ belongs
to some set $S_{w^i}$ and the number of frames accepted as correct
by $s$ is given by $dw^i$. However, to count correctly 
$C\!Acc_s, dw^i$ is too much. One has to withdraw all stations
 accepted by $s$ whose time slots are between $s^i$ and $s$.
This is illustrated in Figure \ref{countAccFork}. 
We introduce auxiliary counters
$d^Aw^k$ and $d^Fw^k$. These counters are set to $0$ when fault $k$ occurs.
Counter $d^Aw^k$ counts how many stations from set $S_{w^k}$ have sent so far,
as counters $dw^k$ do, and counter $d^Fw^k$
counts how many stations from set $S_{w^k}$ were
prevented from sending so far by the clique avoidance mechanism
and moved to the set of non-working stations.
The difference with $dw^k$ is that  these counters  are reset to $0$ each time
a counter $Cp(i)$ reaches $N$ after
fault $k$. Thus $dw^i - \Sigma_{{w'}^k} d^A{w'}^k - \Sigma_{{w'}^k} d^F{w'}^k$,
with $w^i$  a prefix of ${w'}^k$,
 gives exactly 
how many frames between $s$ and $s^{i+1}$  the station 
 $s$ has recognized as correct in the round, and
$dw^i - \Sigma_{{w'}^k} d^A{w'}^k -\Sigma_{{w'}^k} d^F{w'}^k + dw^{i+1} + \cdots + dw^k$ gives
exactly how many frames in total  $s$ has recognized as correct in 
the round, i.e., $C\!Acc_s$. For example, consider again the example illustrated in Figure \ref{countersto4} and a station s from set $S_{0010}$.
Suppose fault 4 occurs  in the first round following fault 2 and station $s$
lies between $s^2$ and  $s^3$, the stations that were emitting when fault 2,
respectively fault 3, occurred. Then:\\
$C\!Acc_s = d_{{00}^2} + d_{{001}^3} + d_{{0010}^4} - d^A_{{0010}^4} -
 d^A_{{0000}^4} - d^F_{{0010}^4} - d^F_{{0000}^4} $.

A similar idea
works for $C\!F\!ail_{s}$. 
This is formally stated in the proposition below. 
\begin{prop}
\label{countkRound1}
We indicate by $w_s$ that the string $w$ refers to an entity 
where $s$ belongs.
\begin{enumerate}
\item
\label{countkRound11}
 Let $s \in S_{w_s^k}$ a station ready to send in the round following  fault
 $k$. 
\begin{enumerate}
\item
If $Cp(1) \leq N$ at the time slot of $s$, then: 
 
$C\!Acc_s = \mid\! W\!\mid - \Sigma_{w'} dw'$, and
$C\!F\!ail_{s} = \Sigma_{w'} dw'$,\\  
where $w'$ must not be a prefix of $w_s^k$.
\item
Let $ 1< i< k$ such that $Cp(i) \geq N$ and $Cp(i+1) < N$ at the time
slot of $s$. Then:

$C\!Acc_s = ( \Sigma_{j=i}^{j=k} dw_s^j) -\Sigma_{{w'}^k} d^A{w'}^k 
-\Sigma_{{w'}^k} d^F{w'}^k $, where $w_s^j$ is a 
 prefix of $w_s^k$ and $w_s^i$ is a prefix of all  ${w'}^k$,  and \\ 
$C\!F\!ail_{s} = (\Sigma_{j=i}^{j=k} \Sigma_{{w}^j\neq w_s^j} d{w}^j) - 
\Sigma_{{w}^k\neq w_s^k} d^A{w}^k   -  \Sigma_{{w}^k\neq w_s^k} d^F{w}^k$ , 
\\ where
 ${w}^k$ must be a suffix of some $w^i\neq w_s^i$. 
\end{enumerate}
\item
\label{countkRound2}
Let $s \in S_w^k$ a station ready to send in the second round 
following  fault $k$. Then:

\noindent
$C\!Acc_s =  dw^k  -d^Fw^k $, and \\
$C\!F\!ail_{s} =  \Sigma_{{w'}^k} d{w'}^k   -  \Sigma_{{w'}^k} d^F{w'}^k$
with ${w'}^k \neq w^k$.
\end{enumerate}
\end{prop}
\begin{proof}
For \ref{countkRound11}, the proof follows what has 
been exposed informally above.
For \ref{countkRound2}, one assumes  counters $dw^k$ are kept as 
there are at the end
of the first round, and that $d^Fw^k$ are reset to $0$ and incremented 
during the second round each time some 
  station  from set $S_{w^k}$ is
prevented from sending. The result follows, since $dw^k$ contains exactly
the number of stations from set $S_{w^k}$ that have sent in the first 
round following fault~$k$.
\end{proof}

\subsection{Complexity issues}

It follows that, for one scenario of the $k$ faults case, the total 
 number of counters needed
is  $\Sigma_{i=1}^{i=k} 2(i+1) \ + k +  2(k+1)$.
There is $\Pi_{i=1}^{i=k} i$ possible scenarios to check.

Using all these counters, an extended automaton similar to the one given in
Figure \ref{pg8}  can be designed and, in theory,
 automatically verified. 
Properties analogous to $P6$ and $P7$ have to be checked to prove
that after the second round following fault $k$, there is only $1$
 clique.
However, in practice, tools that are presently available do not make it possible to handle such a number  of counters already for two faults\footnote{In the case of two faults, we got memory problems both with ALV and LASH.}. Though, a scenario for two faults has been successfully verified  in \cite{BFL04} using their tool FAST after performing further ad hoc abstractions to reduce the number of counters.

\section{Conclusion}
\label{concl}

We have proposed an approach for verifying automatically
a complex algorithm which is industrially relevant.
The complexity of this algorithm is due to its very 
subtle dynamic which is hard to model.
We have shown that this dynamic can be captured 
by means of unbounded (parametric) counter automata.
Even if the verification problem for these 
infinite-state models is undecidable in general, 
there exists many symbolic reachability analysis
techniques and tools which allow to handle such models.

Our approach allows to build a model (counter automaton) 
for the algorithm with an arbitrary number $n$ of stations,
but for a given number $k$ of faults.
We have experimented our approach by verifying in a fully automatic
way the model in the case of one fault, using the ALV tool and the LASH tool.

\medskip

\noindent
{\bf Related Work:}
\cite{bauer} provides a manual proof of the algorithm in the $1$ fault case. 
Theorem \ref{genP1} generalizes this result 
to the case of any number of faults. 
As far as we know, all the existing works on automated proofs 
or verifications of 
the membership algorithm of TTP concern the case of one fault, 
and only symmetric fault occurrences are assumed. 
In our work, we consider the more general framework where 
several faults can occur, and moreover, these faults can be asymmetric. 
In \cite{pfeifer}, a mechanised proof using PVS is provided.
\cite{sri,lakhnech,roscoe} adopt an approach based on combining abstraction
and finite-state model-checking.
\cite{sri} has checked the algorithm for $6$ stations.
\cite{lakhnech,roscoe} consider the parametric verification of $n$
stations; \cite{roscoe} provides an abstraction proved manually
whereas  \cite{lakhnech} uses an automatic abstraction generation technique,
both abstractions leading to
  a finite-state abstraction of the parameterized network.
The abstractions  used in those works seem to be
non-extensible to the case of asymmetric faults and to the $k$ faults case. 
To tackle this more general framework, we provide an abstraction which yields 
a counter automaton and reduce the verification of the algorithm
to the symbolic reachability analysis of the obtained infinite-state
abstract model. Moreover, our abstraction is exact in the sense that it models
faithfully  the emission of frames by stations.

\medskip

\noindent
{\bf Future Work:}
Our future work is to automatize,
for instance using a theorem prover,
the abstraction proof which allows to build
the counter automaton modeling the algorithm.
More generally, an important issue is to design automatic
abstraction techniques allowing to produce infinite-state models
given by extended automata.
Finally, a challenging problem is to design an algorithmic technique allowing
to verify automatically the algorithm by taking into account simultaneously
both of its parameters, i.e., for any number of stations
{\em and for any number of faults}.

\bigskip
{\tt Acknowledgment:}
We thank  anonymous referees for their helpful comments.

\bibliographystyle{acmtrans}
\bibliography{bib}

\end{document}